\documentclass[11pt,final]{article}
\pdfoutput=1
\usepackage{MRnotes}

\newcommand*\pFqskip{8mu}
\catcode`,\active
\newcommand*\pFq{\begingroup
        \catcode`\,\active
        \def ,{\mskip\pFqskip\relax}%
        \dopFq
}
\catcode`\,12
\def\dopFq#1#2#3#4#5{%
        {}_{#1}F_{#2}\biggl(\genfrac..{0pt}{}{#3}{#4};#5\biggr)%
        \endgroup 
        }

\newcommand*\pFqRegskip{8mu}
\catcode`,\active
\newcommand*\pFqReg{\begingroup
        \catcode`\,\active
        \def ,{\mskip\pFqRegskip\relax}%
        \dopFqReg
}
\catcode`\,12
\def\dopFqReg#1#2#3#4#5{%
        {}_{#1}\textbf{F}_{#2}\biggl(\genfrac..{0pt}{}{#3}{#4};#5\biggr)%
        \endgroup 
        }

\catcode`,\active
\newcommand*\IM{\begingroup
        \catcode`\,\active
        \def ,{\mskip\pFqskip\relax}%
        \doIM
}
\catcode`\,12
\def\doIM#1#2#3#4{%
        \mathfrak{T}^{#1}_{_{#2}}\biggl(\genfrac..{0pt}{}{#3}{#4}\biggr)%
        \endgroup 
        }

\newcommand{\ann}{\mathscr{M}}   
\newcommand{\sen}[1]{\varphi_{_{#1}}} 

\newcommand{\wk}{\bar{k}}
\newcommand{\bwt}{\mathfrak{w}}   
\newcommand{\bqt}{\mathfrak{q}}    
\newcommand{\bpt}{\mathfrak{p}} 
\newcommand{\bpts}{\bar{\mathfrak{p}}} 

\newcommand{\ctor}{\zeta}  
\newcommand{\Dz}{\mathbb{D}}  

\newcommand{\In}{\text{\tiny{in}}}     
\newcommand{\Rev}{\text{\tiny{rev}}}  
\newcommand{\Out}{\text{\tiny{out}}}  

\newcommand{\nB}{n_{_B}}   
\newcommand{\Gin}[1]{G^{{#1}}_\In}     
\newcommand{\Gint}[1]{\widetilde{G}^{{#1}}_\In}     
\newcommand{\Kin}[1]{K_{_{#1}}}      
\newcommand{\Grev}[1]{G^{{#1}}_\Rev}   
\newcommand{\Gout}[1]{G^{{#1}}_\Out}   
\newcommand{\Goutt}[1]{\widetilde{G}^{{#1}}_\Out}     
\newcommand{\GR}[1]{G^{{#1}}_\skR}   
\newcommand{\GL}[1]{G^{{#1}}_\skL}   
\newcommand{\Gbulk}[1]{\mathcal{G}^{{#1}}_\text{\tiny{bb}}}   
\newcommand{\gfn}[1]{\mathfrak{G}_{_{#1}}}

\newcommand{\Krev}[1]{\overline{K}_{_{#1}}}    

\newcommand{\skR}{\text{\tiny R}}
\newcommand{\skL}{\text{\tiny L}}

\newcommand{\JF}{J_{_\text{F}}}
\newcommand{\JP}{J_{_\text{P}}}
\newcommand{\JFn}{\widehat{J}_{_\text{F}}}
\newcommand{\JPn}{\widehat{J}_{_\text{P}}}

\newcommand{\snMar}{ \breve{\Phi}}
\newcommand{\snMarP}{ \breve{\Phi}_{_\text{P}}}
\newcommand{\snMarF}{ \breve{\Phi}_{_\text{F}}}

\title{Holographic open quantum systems: Toy models and analytic properties of thermal correlators}
\author[a]{R. Loganayagam,}  
\author[b]{ Mukund Rangamani,}
\author[b]{ Julio Virrueta}

\affiliation[a]{
    International Centre for Theoretical Sciences (ICTS-TIFR), \\ 
    Tata Institute of Fundamental Research, Shivakote, Hesaraghatta, Bangalore 560089, India.}
 
\affiliation[b]{
    Center for Quantum Mathematics and Physics (QMAP)\\
    Department of Physics \& Astronomy, University of California, Davis, CA 95616 USA}
\emailAdd{nayagam@icts.res.in}
\emailAdd{mukund@physics.ucdavis.edu}
\emailAdd{jvirrueta@ucdavis.edu}

\abstract{ 
We present a unified picture of  open quantum systems, the theory of a system probing a noisy thermal environment, distilling lessons learnt from previous holographic analyses. Our treatment is applicable both when the system is coupled to short-lived (Markovian), and long-lived (non-Markovian) environmental degrees of freedom. The thermal environment is modeled using an asymptotically AdS black hole, and the systems of interest are simple probe field theories. The effective stochastic dynamics of the system is governed by real-time thermal correlators, which we compute using the gravitational Schwinger-Keldysh (grSK) geometry. We describe the structure of arbitrary tree-level contact and exchange Witten diagrams in the grSK geometry. In particular, we argue, that all such diagrams reduce to integrals supported on a single copy of the exterior of the black hole. The integrand is obtained as a multiple discontinuity of a function comprising ingoing boundary-bulk propagators, monodromy functions which appear as radial Boltzmann weights, and vertex factors.  These results allow us to deduce the analytic structure of   real-time thermal $n$-point functions in holographic CFTs. We illustrate the general statements by a two-dimensional toy model, dual to fields in the BTZ background, which we argue captures many of the essential features of generic open  holographic QFTs.
}

\begin{document}
\maketitle

 
\section{Introduction}
\label{sec:intro}

The basic principles underlying the effective dynamics of open quantum systems are well-understood, cf., \cite{Feynman:1963fq}. However, actual progress in the field theoretic context has been limited, in part, for technical reasons. In weakly coupled systems there is no obvious scale separation (environmental dynamics in such situations is slow) leading to long-time effects in the effective description. Motivated by this, \cite{Jana:2020vyx} argued for using holographic field theories as thermal environments to model open quantum dynamics. We shall further explore this framework and provide an illustration of the general lessons learned from the holographic analysis in the context of an analytically tractable low-dimensional toy-model.

Our set-up is the following: a finite temperature holographic field theory provides a strongly correlated environment, which we probe by a simple quantum system, a free scalar field coupled to a gauge invariant single-trace operator. The effective field theory of the scalar probe depends on the  real-time thermal correlation functions of the holographic theory. These are computed on the complex gravitational Schwinger-Keldysh geometry introduced in \cite{Glorioso:2018mmw} (important earlier work on holographic computations of Schwinger-Keldysh correlators includes \cite{Son:2002sd,Herzog:2002pc,Skenderis:2008dg,vanRees:2009rw}). 

The authors of \cite{Jana:2020vyx} focused on the case where the single trace operator was a conformal primary. These operators typically relax within a thermal timescale. This fast relaxation was the motivating factor behind the choice of holographic environments as an ideal test-bed for analyzing open quantum dynamics. However, holographic environments also have slowly relaxing modes, generically corresponding to components of conserved current operators.  They lead to hydrodynamic behaviour of the holographic plasma. Our aim is to give a unified treatment which subsumes both long-lived and short-lived operators, which following \cite{Ghosh:2020lel} we refer to as non-Markovian and Markovian, respectively.  

For example, the energy-momentum tensor of a thermal system has both such modes. Tensor polarizations are Markovian (short-lived), while transverse vectors describing momentum diffusion, and longitudinal scalar polarizations capturing the phonon mode are non-Markovian (long-lived).\footnote{
      The polarization decomposition is with respect to the little group of rotations in the space orthogonal to a fixed vector (the direction of motion).
} 
In \cite{Ghosh:2020lel,He:2022jnc} the authors constructed the Gaussian effective action for these modes.\footnote{
    An analogous analysis for finite charge density environments was carried out in \cite{He:2021jna,He:2022deg}.}  
The hydrodynamic behaviour of course has been  well-known for a long time at the level of dispersion relations \cite{Policastro:2002se,Policastro:2002tn}.

At this stage we should clarify our terminology a bit. Conventionally, in the open quantum system literature, non-Markovian dynamics refers to situations where there is long-time temporal correlation leading to memory effects. Typically, this arises because the environment that is being integrated out has such long-time dynamics. In our examples, we are explicitly coupling our probe systems to operators of the thermal environment whose correlation functions exhibit long-distance and large-time effects. Furthermore, these arise because of the presence of low-lying Goldstone type collective modes. We will therefore adapt the definition non-Markovian systems to characterize modes of the environment itself. In particular, we refer to modes of the environment that have large infra-red effects as non-Markovian. As noted above, a single (gauge invariant) operator of the environment might itself have contained Markovian and non-Markovian components, which one would have to disambiguate carefully to obtain a local open effective action.

The holographic analysis was aided  by repackaging different polarizations of currents  in terms of diffeomorphism invariant combination of gravitational perturbations \cite{Kodama:2003jz}. These combinations end up being non-minimally coupled scalar fields in the background with  their kinetic terms modulated by an auxiliary dilaton. For massless fields like the graviton, the Markovian versus non-Markovian nature can be deduced from the asymptotic behaviour of the dilaton.\footnote{
      Markovian fields were found to be repelled from the asymptotic boundary, while the non-Markovian fields face no action penalty for hovering about the asymptopia. 
} It was useful to characterize the distinction using a single parameter, the  Markovianity index $\ann$, as in the aforementioned works. As explained in \cite{He:2022jnc}, for massless bulk fields,  $\ann \geq 1$ corresponds to Markovian operators, $\ann \leq-1$ to non-Markovian operators, and $\ann \in (-1,1)$ could be either depending on the boundary conditions.\footnote{
   A secondary feature of the non-minimal coupling, which will not be of interest in the present work, is a mass (or potential) term that is modulated along the radial direction.}

For energy-momentum dynamics in a $d$-dimensional conformal plasma,  gravitational dynamics breaks up into a set of $\frac{d(d-3)}{2}$ Markovian modes with $\ann=d-1$ $(d-2)$ non-Markovian modes with $\ann =1-d$, and a single scalar mode with $\ann = 3-d$. These modes can be characterized by their transformation under the $SO(d-2)$ little group, corresponding to rotations about a fixed spatial vector (taken to be the direction of spatial momentum).  The Markovian modes correspond to the transverse traceless spin-2 polarizations of the energy-momentum tensor (roughly $T^{ij}$), the $d-2$ non-Markovian mode are the transverse vector polarizations (roughly $T^{0i}$), and the single longitudinal mode is the energy density (roughly $T^{00}$).
The quadratic action for these modes has been derived from Einstein-Hilbert in \cite{Ghosh:2020lel,He:2022jnc} (and for the Einstein-Maxwell action in \cite{He:2021jna,He:2022deg}). Higher order graviton vertices are necessary to extract the non-Gaussian influence functionals.  Rather than undertake this exercise (which is straightforward but somewhat technical) we identify a simple toy model where we can exhibit all the essential features with the added benefit of analytic tractability.

Our discussion broadly comprises two parts. We  first present an abstract  model which distills the analysis of gravitational fluctuations. Here
we write down an interacting theory of designer scalar fields, analyze the quantization conditions, and give the general rules for computing boundary observables, generalizing  \cite{Jana:2020vyx}. A new element in our presentation is an evaluation of bulk exchange diagrams for real-time Schwinger-Keldysh correlators. In fact, we will argue that a general tree-level Witten diagram can be computed given the knowledge of the  ingoing boundary-bulk propagator (similar observations have been made earlier in \cite{Arnold:2011hp}). We furthermore argue that the general diagram can be evaluated on a single copy of the exterior of the black hole, with the integrand given as a multiple discontinuity \cite{Loganayagam:2022xyz}. We illustrate this explicitly for low-point correlators, giving expressions for four-point Schwinger-Keldysh correlators in terms of a single master bulk integral. 

In the second part, we specify this model to the particular case of the BTZ geometry.  This has the advantage that the boundary-bulk propagator may be obtained analytically in closed form in terms of hypergeometric functions.\footnote{
   This is also the case for a three-dimensional system with broken spatial translations analyzed in \cite{Davison:2014lua}. The analytic structure of the retarded boundary correlators (for probes that are insensitive to the broken translations) in the two models is identical.}
We will furthermore highlight the fact that the designer scalars can have long-lived diffusive quasinormal modes even in this low dimension. Recall that in a two-dimensional thermal CFT energy-momentum dynamics is fixed by the Virasoro symmetry; the system only has left and right moving modes, and thus no hydrodynamic behaviour. While the designer scalars are not components of the conserved current (bulk graviton dynamics in \AdS{3} is trivial), the model captures the essential features of the higher dimensional physics in a tractable setting.\footnote{ 
   In two spacetime dimensions, long-lived diffusive dynamics will be associated with strong infra-red quantum (loop) fluctuations. As we wish to illustrate the features that appear in the physical higher dimensional case, we will not attempt to read too much into the dynamics of diffusion in low-dimensional systems. }

With benefit of hindsight from our model, we also describe the analytic structure of higher-point Schwinger-Keldysh thermal correlators. We argue that a generic $n$-point thermal correlator will be meromorphic, with quasinormal or anti-quasinormal poles depending on whether the operator insertion is retarded or advanced. The poles appear in the frequencies corresponding to the external operator insertions, and also in the internal operator exchanges. The analysis is facilitated by use of thermally adapted advanced retarded basis introduced in \cite{Chaudhuri:2018ymp}, which judiciously factors out the thermal statistical factors associated with the KMS constraints on real-time correlators. In addition to these quasinormal poles, we also find that exchanged operator frequencies can have Mastubara poles outside their domain  of analyticity. These occur at discrete multiples of $2\pi T$, with retarded (advanced) operators potentially supporting such in the lower (upper) half-plane.

The outline of the paper is as follows: In \cref{sec:designer} we introduce the general class of models that are of interest and work out the appropriate generalization of the holographic GKPW dictionary. Specifically, in \cref{sec:Wittendiag} we outline the computation of bulk exchange diagrams in the grSK geometry. We also take the opportunity to explain the analytic structure of real-time holographic thermal correlators in \cref{sec:analcorr}. In particular, we argue that even higher-point functions only have quasinormal poles (and potentially Matsubara poles in exchanged momenta).

\cref{sec:BTZtoy} is devoted to our three-dimensional toy model. We first study designer fields in the BTZ background in \cref{sec:setup}, obtaining analytic expressions for the bulk Green's functions and the boundary two-point correlators \cref{sec:gaussian}. The analysis of the correlators reveals interesting linear (mode) instability domains in our model (\cref{sec:astab}). In \cref{sec:nongauss} we compute three and four-point functions. While the explicit expressions are somewhat complicated, we express them optimally to extract physical lessons for open quantum systems, which are summarized \cref{sec:phylessons}. We conclude with a broader discussion and potential generalizations in \cref{sec:discuss}. Some technical details for computing exchange diagrams in the grsK geometry can be found in \cref{sec:exchangechecks}.

\section{Designer fields in grSK spacetime}
\label{sec:designer}

The problem we want to consider is that of a simple scalar field $\Psi$ coupled to a holographic thermal field theory. Let $\mathcal{O}_a$ be a set of single trace operators of the holographic theory. We will be interested in both short-lived (Markovian) and long-lived (non-Markovian) operators. For the present, we are going to elide over the tensor indices of the operator. The reader might find it helpful to view $\mathcal{O}_a$ as specific polarizations of a single tensor operator. As discussed in \cref{sec:intro} this is the case for the energy-momentum tensor, with the different components having Markovian, or non-Markovian characteristics.

The combined dynamics is specified by the action of the schematic form
\begin{equation}\label{eq:bulkmodel}
S = S_\text{hol} + S[\Psi] + \int \, d^dx\,  \sum_a\, \alpha_a[\Psi] \,\mathcal{O}_a\,.
\end{equation}  
We have chosen to represent the coupling between the holographic theory and our system field $\Psi$ using a functional $\alpha_a[\Psi]$ to keep track of the dependence on tensor components, polarizations, etc.\footnote{
   We use the index $a$ to refer to the species of the designer field.
}
Since the coupling is directly to the gauge invariant operators, it follows that the functionals $\alpha_a[\Psi]$ are simply sources for $\mathcal{O}_a$. Therefore, insofar as the effective field theory of $\Psi$ is concerned, we first need the data of the real-time correlation functions of $\mathcal{O}_a$, which as noted above are computed holographically in the grSK geometry. This will allow us to write down the leading order (in amplitudes) the effective stochastic theory for $\Psi$, cf., \cite{Jana:2020vyx}.

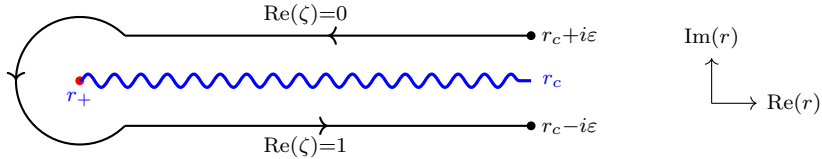
\begin{figure}[h!]
\begin{center}
\begin{tikzpicture}[scale=0.6]
\draw[thick,color=red,fill=red] (-5,0) circle (0.45ex);
\draw[thick,color=black,fill=black] (5,1) circle (0.45ex);
\draw[thick,color=black,fill=black] (5,-1) circle (0.45ex);
\draw[very thick,snake it, color=blue] (-5,0) node [below] {$\scriptstyle{r_+}$} -- (5,0) node [right] {$\scriptstyle{r_c}$};
\draw[thick,color=black, ->-] (5,1)  node [right] {$\scriptstyle{r_c+i\varepsilon}$} -- (0,1) node [above] {$\scriptstyle{\Re(\ctor) =0}$} -- (-4,1);
\draw[thick,color=black,->-] (-4,-1) -- (0,-1) node [below] {$\scriptstyle{\Re(\ctor) =1}$} -- (5,-1) node [right] {$\scriptstyle{r_c-i\varepsilon}$};
\draw[thick,color=black,->-] (-4,1) arc (45:315:1.414);
\draw[thin, color=black,  ->] (9,-0.5) -- (9,0.5) node [above] {$\scriptstyle{\Im(r)}$};
\draw[thin, color=black,  ->] (9,-0.5) -- (10,-0.5) node [right] {$\scriptstyle{\Re(r)}$};  
\end{tikzpicture}
\caption{ The contour picked by the grSK geometry in the complexified $r$ plane with the ingoing time coordinate $v$ kept fixed.  The contour encircles the branch point at the horizon $r=r_+$ counter-clockwise starting from left SK boundary (denoted L) at $r_c+i\varepsilon$ and ending up at the right boundary  (denoted R) at $r_c-i\varepsilon$. }
\label{fig:mockt}
\end{center}
\end{figure}

The grSK spacetime is a two-sheeted geometry characterized by a complex line element. In the conventions of \cite{Jana:2020vyx}, a stationary asymptotically AdS black hole geometry in ingoing Eddington-Finkelstein coordinates is extended to the complex domain. Consider
\begin{equation}\label{eq:sadsct}
ds^2 = -r^2\, f(r) \, dv^2 +  i\, \beta\, r^2 \, f(r)\,  dv\, d\ctor + r^2\, d\vb{x}^2 \,,
\end{equation}  
parameterized by the  mock tortoise coordinate $\ctor$. Assuming the  emblackening function $f(r)$ to have a simple zero at $r=r_+$, the location of the horizon, $\ctor$ is defined  by 
\begin{equation}\label{eq:ctordef}
\frac{dr}{d\ctor} = \frac{i\,\beta}{2} \, r^2\, f(r) \,, \qquad \beta = \frac{4\pi}{f'(r_+)} \,,
\end{equation}  
subject to the following boundary conditions at the cut-off surface $r=r_c$
\begin{equation}\label{eq:ctorbc}
\ctor(r_c+i\,\varepsilon) = 0 \,, \qquad \ctor(r_c-i\,\varepsilon) = 1\,.
\end{equation}  
This geometry has two asymptotic boundaries which are the backward and forward segments of the boundary Schwinger-Keldysh time contour. The bulk spacetime can be simply characterized by a contour definition for $\ctor$ which encircles the horizon at $r=r_+$, cf., \cref{fig:mockt}.

The operators $\mathcal{O}_a$ are dual to fields $\sen{a}$ in the bulk geometry. These fields are not minimally coupled but have a modulation of their gravitational interaction by an auxiliary dilaton $\chi_a(r)$. We consider a number of these operators having different bulk modulations interacting via a cubic coupling, a bulk three-point vertex for the fields $\sen{a}$. Thus, the action for these fields takes the form
\begin{equation}\label{eq:Sdesign}
\begin{split}
S[\sen{a}] 
&=  \, \int\, d^{d+1}x\, \sqrt{-g} \left[ -\frac{1}{2}
   \sum_a \; e^{\chi_a(r)} \, \left(\nabla_A \sen{a} \nabla^A\sen{a} + m_a^2 \, \sen{a}^2\right) - \sum_i \, \lambda_i\, e^{\chi_{\lambda_i}(r)} \sen{a_i} \, \sen{b_i}\, \sen{c_i} \right] \\
&\qquad
 + S_\text{bdy} + S_\text{ct}\,.
\end{split}
\end{equation}  

The asymptotic behaviour of the dilaton has implications for the boundary conditions, and in certain cases also for the nature of the operator $\mathcal{O}_a$.  We characterize the asymptotic fall-off of this auxiliary dilaton by an index $\ann_a$
\begin{equation}\label{eq:chiasym}
\lim_{r\to \infty} \, e^{\chi_a} = 
   \left( \frac{r_+}{r}\right)^{\ann_a-d+1} \,.
\end{equation}  
We have chosen a minimally coupled scalar field to have index $\ann=d-1$. In addition, we have allowed ourselves the freedom for the vertex function to also have a radial modulation through another auxiliary dilaton $\chi_\lambda(r)$, which is uncorrelated to the modulation in the kinetic terms. The boundary terms in $S_\text{bdy}$ include the appropriate boundary terms to ensure the stationarity of the action (discussed below), while the counterterms $S_\text{ct}$ provide the regularization of the bulk computations.  

The motivation for this model comes from the structure of the gravitational fluctuations around a neutral AdS black hole decomposed in gauge invariant variables. As explained in \cite{Ghosh:2020lel,He:2022jnc} (following the original derivation of \cite{Kodama:2003jz}) the energy-momentum tensor can be decomposed into polarizations based on the little group orthogonal to a chosen spatial momentum $\vb{k}$ in $SO(d-2)$ irreps. Transverse traceless tensor polarization are Markovian with $\ann = d-1$, transverse vectors and the longitudinal scalar  are non-Markovian with indices $\ann = 1-d$ and $\ann = 3-d$, respectively. If we decompose the Einstein-Hilbert action in terms of the gauge invariant combinations of gravitational fluctuations the resulting action truncated to quadratic order will be of the form given in \eqref{eq:Sdesign}. The quadratic action for the fields can be found in the aforementioned references, but the cubic vertices have not yet been evaluated directly. We have simply distilled the essential features into the simple model above.\footnote{ 
   Note that the cubic couplings of gravitons will also involve derivatives.
    In this paper we only focus on non-derivative interaction terms. As we note later, derivative interactions lead to additional subtleties with localized contributions at the horizon \cite{Chakrabarty:2019aeu}.\label{fn:derint}
}
In general the dilatonic couplings in the kinetic terms $\chi_a(r)$ are functions of the radial coordinate -- either simple power laws, or some functions constructed from the background metric data. An exception is energy density operator, the scalar polarization of the gravitons, where the modulation also depends on the spatial momentum, see \cite{He:2022jnc,He:2022deg}.

For the present, we will focus on this model and explain the general features for the computation of the real-time correlation functions of $\mathcal{O}_a$, which can then be translated into an open stochastic effective action for the field $\Psi$. Much of this has already been explained in \cite{Jana:2020vyx} and \cite{Ghosh:2020lel}, so we will be brief, and only highlight the salient features. The main new ingredient not present in these works is a discussion of the bulk-bulk propagator, the computation of exchange diagrams in the bulk, 
and a discussion of the analytic structure of the correlators. 

\subsection{Scalar propagation in grSK geometries}
\label{sec:sprops}

The essential data we need for setting up the computation of real-time thermal correlators from the grSK geometry is the boundary-bulk propagator with ingoing boundary conditions at the horizon. This will suffice to construct the solution of the homogeneous wave equation on the grSK geometry with suitable boundary conditions for the field, and will also determine the bulk-bulk propagator.

A scalar field $\sen{}$ with dilaton $\chi(r)$ satisfies the free wave equation\footnote{ 
   We drop the species label for the present and reinstate it later when it becomes necessary to distinguish between the operators.}
\begin{equation}\label{eq:homeqn}
\left(\mathfrak{D} + m^2 \right) \sen{}= 0\,, \qquad \mathfrak{D} \equiv -\frac{1}{\sqrt{-g}} \partial_A\left(e^{\chi}\, \sqrt{-g}\, g^{AB}\, \partial_B\right).
\end{equation}	
Using the asymptotic behaviour of the dilaton \eqref{eq:chiasym} we determine the linearly independent solutions near the AdS boundary
\begin{equation}\label{eq:phiasym}
\sen{} = c_1\, \frac{1}{r^{\Delta}} + c_2 \, \frac{1}{r^{1+\ann-\Delta}}\,,
\end{equation}	
where we defined `conformal dimensions' $\Delta$ to satisfy\footnote{
   We are not demanding that $\varphi$ be dual to a boundary conformal primary. We view it as an effective field repackaging components of primaries, and therefore, the fall-offs a-priori do not map to conformal dimensions of boundary operators. For example, this is the case  for the scalar and transverse vector components of the energy-momentum tensor. They  correspond to designer scalars (after stripping of polarization labels) of index $\ann = 3-d$ and $\ann =1-d$, respectively, while the energy-momentum tensor itself is a primary of dimension $d$.
    However, since in the bulk description, we only have an effective field, we will be sloppy and refer to $\Delta$ as the conformal dimension, with this caveat implicit in the sequel. 
}
\begin{equation}\label{eq:cdimdef}
\Delta\,(\Delta - 1-\ann) = m^2 \,.
\end{equation}	
The fall-offs reduce to the familiar expressions for minimally coupled fields whence $\ann = d-1$. Note that with our choice for the dilatonic modulation \eqref{eq:chiasym} we effectively can think of the scalars as propagating on an \AdS{} spacetime with effective dimension $d_\text{eff} = 1+ \ann$ (similar features were observed earlier in a different context in \cite{Caldarelli:2013aaa}).

We now have to decide how to quantize the fields given these fall-offs. As such, we require the masses to satisfy $m^2 \geq -\frac{(\ann+1)^2}{4}$, the modified Breitenlohner-Freedman bound for the dimension $\Delta$ to be real. This reality condition ensures that $\mathfrak{D}$ is self-adjoint. We will assume this to be case and focus on real $\Delta$ for the reminder of our analysis.

The possible boundary conditions now depend on choice of $\ann$ and $\Delta$. We do not want to enforce that the $\Delta$ satisfy the unitarity bound  relevant for regular conformal dimensions. Since the operator $\mathcal{O}$ is to be viewed as a component of a gauge invariant operator satisfying unitarity, it by itself does not need to satisfy the usual restrictions. So apart from requiring $\Delta \in \mathbb{R}$ we won't prejudice the model with further restrictions. Later, in our two-dimensional example we will see a constraint from mode stability which cuts-off a part of the $(\ann,\Delta)$ space.

To proceed, we also note that conjugate momentum  $\pi$ behaves as 
\begin{equation}\label{eq:conjmom}
\begin{split}
    \pi &\equiv  -\sqrt{-g}\, e^{\chi}\, g^{rA}\, \partial_A\, \sen{} \,, \\ 
    \lim_{r\to \infty}\, \pi &= c_1\, \Delta\, \frac{1}{r^{\Delta-1-\ann}} -c_ 2 (\Delta -1-\ann) \,r^{\Delta} \,.
\end{split}
\end{equation}	
We can therefore postulate the following boundary conditions for computing the generating functional of correlators:

\begin{itemize}[wide,left=0pt]
\item For $m^2=0$ we have the modes falling off as a constant and $r^{-1-\ann}$, respectively. This singles out $\ann=-1$ as the separatrix between two possibilities. For $\ann > -1$ and $\ann <-1$ we impose Dirichlet and Neumann boundary conditions, respectively. In the window $\ann \in [-1,1]$ we are free to impose either (or even mixed boundary conditions) based on the choice of boundary terms. This was the situation encountered for the designer scalars in higher dimensions \cite{Ghosh:2020lel,He:2022jnc} (and perhaps most physical). 
\item  When $m^2 \neq 0$ the fall-offs are separated across the locus $\Delta = \frac{1+\ann}{2}$, irrespective of the magnitude and sign of $\ann_a$.   We should impose Dirichlet boundary conditions when $\Delta \geq \frac{1 + \ann}{2}$, identifying the mode falling off as $r^{\Delta -1 - \ann_a}$ as the non-normalizable mode. Following the usual rules of quantization in AdS/CFT we will therefore define the source of the operator $\mathcal{O}$ in this case to be
\begin{equation}\label{eq:sourceJ}
    J(v,\vb{x}) = \lim_{r\to \infty}\, r^{1+\ann - \Delta}\, \sen{} \,.
\end{equation}	
When  $\Delta < \frac{1+\ann}{2}$ we should instead impose Neumann boundary conditions
\begin{equation}\label{eq:sourceJalt}
    \widehat{J}(v,\vb{x}) = \frac{1}{\Delta}\lim_{r\to \infty}\, r^{\Delta-1-\ann}\, \pi \,.
\end{equation} 
\item We need to include suitable boundary terms in order to ensure that the variational principle is upheld. For the Dirichlet boundary condition $S_\text{bdy,D} =0$ since the variation of the bulk action gives a boundary term proportional to $\pi\,\delta \sen{}$. On the other hand, to impose the Neumann boundary condition we need to include a boundary term $S_\text{bdy,N} = -\int d^2 x\, \pi\, \sen{}$.  

\end{itemize}

The Markovian or non-Markovian nature of the operator $\mathcal{O}$ is not a-priori dictated from the boundary conditions alone. We define an operator to be Markovian if the boundary retarded Green's function is analytic in Fourier domain at low momenta and frequencies. Otherwise, it will be designated to be non-Markovian. In particular, non-Markovian operators will have thermal correlators which have quasinormal poles with dispersion relation $\omega(k) \to 0$ as $k \to 0$. 

Markovian operators encountered in the literature  are dual to massless fields ($m^2=0$) with  $\ann >-1$  quantized with Dirichlet boundary conditions, while non-Markovian operators correspond to fields with $\ann <1$ (and $m^2=0$) quantized with Neumann boundary conditions \cite{Ghosh:2020lel,He:2022jnc}.  
However,  is worth emphasizing that the choice of boundary conditions is independent of whether the operator has long-lived or short-lived characteristics. Indeed, we shall see examples of this below, for once we have a mass term, there is a possibility of encountering non-Markovian behaviour with either Dirichlet or Neumann boundary conditions.

\paragraph{Boundary-bulk propagator:}
With the boundary conditions specified, we can now give the prescription for computing correlation functions. The only piece of data we need is the ingoing boundary-bulk propagator, $\Gin{}(\ctor,v,\vb{x})$. It will be convenient to work in Fourier domain where\footnote{ 
   Fourier conventions: We define the $d$-momentum $k^\mu = (\omega,\vb{k})$, but refrain from indicating the Lorentz index. Spatial momentum magnitude will be denoted by  $\abs{\vb{k}}$. The frequency reversed $d$-momentum is singled out by a bar decoration: $\wk^\mu = (-\omega,\vb{k})$. Finally,  the Fourier transforms is defined as  
   \[
      \mathfrak{f}(\ctor,v,\vb{x})
      =
      \int \, \frac{d\omega}{2\pi}\, \frac{d^{d-1}\vb{k}}{(2\pi)^{d-1}} \,
         \mathfrak{f} (\ctor,\omega,\vb{k})\,
         e^{-i\,\omega \,v + i\,\vb{k}\cdot \vb{x}} 
      \equiv 
      \int_k\, \mathfrak{f}_k \, e^{ikx} \,.
   \]
   We also abbreviate the momentum integrals by $\int_k$ as indicated above.
}
\begin{equation}\label{eq:Gin}
\lim_{r\to r_c} \, r^{1+\ann-\Delta}\,\Gin{}(\ctor,k) = 1\,, \qquad 
\lim_{r\to r_+} \, \Gin{}(\ctor,k) = \textrm{regular} \,.
\end{equation} 
We have given the definition for Dirichlet boundary conditions, which can straightforwardly be generalized to the Neumann case.\footnote{ 
   For the Neumann boundary condition we would require the source, determined by the conjugate momentum, to be suitably normalized at the boundary and regular at the horizon. }

It suffices to determine the ingoing Green's function $\Gin{}$ on one of the sheets of the grSK geometry. It is regular as $\ctor$ jumps across the sheets for it satisfies
\begin{equation}\label{eq:Ginperiod}
\Gin{}(\ctor +1,k) = \Gin{}(\ctor,k)\,.
\end{equation} 
Given the ingoing boundary-bulk Green's function we can obtain the outgoing Green's function by conjugation  \cite{Jana:2020vyx}
\begin{equation}\label{eq:Gout}
 \Gout{}(\ctor,k) 
 = e^{-\beta\omega\,\ctor}\, \Gin{}(\ctor,\wk)\,,
\end{equation} 
where $\wk^\mu$ is the $d$-momentum with frequency reversed, $\wk^\mu \equiv (-\omega, \vb{k})$. The exponential factor $e^{\beta\omega \zeta}$, which arises as the monodromy around the horizon along the grSK contour, acts as a \emph{radial Boltzmann weight}. We will adapt this terminology in the sequel.  

It is also useful to record the time-reversed propagator 
\begin{equation}\label{eq:Grev}
\Grev{}(\ctor,k) = \Gin{}(\ctor,\wk)   = \Gin{}(\ctor,-\omega,\vb{k}) \,.
\end{equation} 

The general solution to the homogeneous wave equation \eqref{eq:homeqn} with sources $J_\skR$ and $J_\skL$ prescribed on the two boundaries (R and L, respectively) of the grSK geometry is then
\begin{equation}\label{eq:homsoln}
\sen{}(\ctor,k)
=
   \Gin{}(\ctor,k) 
      \bigg( (1+\nB) \, J_\skR - \nB\, J_\skL\bigg) 
   - \Grev{}(\ctor,k)  \,
       e^{\beta \,\omega \, (1-\ctor)}\, \nB \, \bigg(J_\skR -J_\skL\bigg) \,. 
\end{equation} 
Here $\nB$ is the Bose-Einstein statistical factor
\begin{equation}\label{eq:BE}
\nB(\omega) = \frac{1}{e^{\beta\omega} -1} \,.
\end{equation} 
The combination of sources appearing above can be used to define a variant of the  retarded-advanced basis introduced in \cite{Chaudhuri:2018ymp}
\begin{equation}\label{eq:retadv}
\JF = - \bigg( (1+\nB) \, J_\skR - \nB\, J_\skL\bigg) \,, \qquad 
\JP = - \nB \, \bigg(J_\skR -J_\skL\bigg) \,.
\end{equation} 
As noted in \cref{sec:intro} an advantage of this basis is that it factorizes out (for the external operators) the statistical factors $\nB$ which arise from the KMS conditions on thermal correlators. 

It then follows that the solution for the scalar field on the grSK geometry is given as
\begin{equation}\label{eq:homsolFP}
\sen{}(\ctor,v,\vb{x}) = - \Gin{}(\ctor,v,\vb{x}) \, \JF + \Grev{}(\ctor,v,\vb{x}) \, e^{\beta\omega\,(1-\ctor)} \, \JP \,.
\end{equation} 
We will use this form of the solution in what follows since the Schwinger-Keldysh and KMS conditions imply that any correlator with all operators solely of either the P or F-type vanishes. The only non-vanishing correlators are of the mixed type \cite{Chaudhuri:2018ymp}. 

From the boundary-bulk ingoing propagator we can obtain the boundary retarded Green's function $\Kin{}(k)$, using the fact that it is given by the asymptotic value of the field and the conjugate momentum, viz., 
$\Kin{}(k) = \lim_{r\to \infty} \, \Gin{}(\ctor,k) \, \pi(\ctor,k) $. This was justified in the grSK geometry in \cite{Jana:2020vyx}. 

\paragraph{Bulk-bulk propagator:}
The other piece of data we require is the bulk-bulk propagator 
$\Gbulk{}(X,X')$, where $X = \{\ctor,v,\vb{x}\}$ for brevity. This satisfies the wave equation with a delta function source 
\begin{equation}\label{eq:inhomeqn}
\left(\mathfrak{D}_{a} + m^2 \right) \Gbulk{}(X,X')= 
   \frac{1}{\sqrt{-g}\, e^{\chi}} \, \delta^{(d+1)}(X-X')\,.
\end{equation} 
One can obtain this Green's function using the variation of parameters trick, which  exploits the solutions of the homogeneous wave equation 
\eqref{eq:homeqn}. To write this efficiently, let us introduce a linear combination of boundary-bulk Green's functions that are normalizable at one or the other boundary of the grSK geometry. Denoting these as $\GR{}(\ctor,k)$ and $\GL{}(\ctor,k)$, respectively, we have 
\begin{equation}\label{eq:GLRdef}
\begin{split}
\GR{}(\ctor,k)
&=
        e^{\beta\,\omega}\, \nB \left(\Gin{}(\ctor,k) - \Gout{}(\ctor,k)\right) ,
    \\ 
\GL{}(\ctor,k)
&=
     -\nB \left(\Gin{}(\ctor,k) - e^{\beta\omega}\, \Gout{}(\ctor,k) \right) .
\end{split}
\end{equation}  
The function $\GR{}$ has a source on the right boundary ($\ctor =1$), while $\GL{}$ has a source on the left boundary ($\ctor =0$), and they are respectively normalizable at the other end viz.,
\begin{equation}\label{eq:GLRbcs}
\lim_{\ctor \to 0}\, \{\GR{}, \GL{}\} = \{0,1\}  \,, \qquad 
\lim_{\ctor \to 1}\, \{\GR{}, \GL{}\} = \{1,0\}  \,.
\end{equation}  

The bulk-bulk propagator can be seen to be given by  
\begin{equation}\label{eq:Gbulk}
\begin{split}
\Gbulk{ }(\ctor,\ctor';k)
&=  
   \mathscr{N}(k)\;  e^{\beta\omega\,\ctor'} 
   \bigg[\Theta(\ctor - \ctor') \, \GL{ }(\ctor, k) \, \GR{ }(\ctor', k) 
   +\Theta(\ctor'- \ctor) \, \GL{ }(\ctor', k)\, \GR{ }(\ctor, k)  \bigg] \,.
\end{split}
\end{equation} 
Here $\Theta(\ctor-\ctor')$ is a contour ordered step function along the contour depicted in \cref{fig:mockt}. The prefactor can be obtained from the Wronskian of the two linearly independent solutions to the homogeneous equation which we have taken to be the left and right normalizable boundary-bulk propagators. We have separated this out into a normalization factor
$\mathscr{N}(k)$ and a piece that depends on the radial position of the delta function source. Of these, the  $e^{\beta\omega\,\ctor'}$ factor will be  crucial -- it ensures that the observables satisfy the Schwinger-Keldysh and KMS constraints. We can furthermore show that the normalization factor  $\mathscr{N}(k)$ can be determined in terms of the boundary retarded Green's function $K(k)$ as
\begin{equation}\label{eq:Nkgeneric}
\mathscr{N}(k) = \frac{1}{\nB+1} \, \frac{1}{K(k) - K(\wk)}\,.
\end{equation}  

To derive these statements, first, recall that the overall normalization of the bulk-bulk propagator is given in terms of the Wronskian between $\GR{}$ and $\GL{}$. This in turn is related to the Wronskian between $\Gin{}$ and $\Gout{}$ through
\begin{equation}\label{eq:WRLio}
\text{Wr}(\GR{},\GL{}) =  (\nB+1)\, \text{Wr}(\Gin{},\Gout{}) \,.
\end{equation}  
The latter can be computed as follows:
\begin{equation}\label{eq:Wioderive}
\begin{split}
 \text{Wr}(\Gin{},\Gout{}) 
 &= 
     e^{\chi}\, \left(\Gin{}(k)\, D_\ctor^+ \Gout{}(k) - \Gout{}(k)\, D_\ctor^+ \Gin{}(k)\right) , \\
 &= 
     e^{\chi}\, e^{-\beta\omega\ctor} \left(\Gin{}(k)\,  D_\ctor^- \Gin{}(\wk) - \Gin{}(\wk)\, D_\ctor^+ \Gin{}(k)\right), \\
 &= - e^{-\beta\omega\ctor}\, \left(\Gin{}(k)\, \pi(\wk) - \pi(k)\, \Gin{}(\wk) \right) , \\
 & =  e^{-\beta\omega\ctor}\, \left(\Kin{}(k)-\Kin{}(\wk)\right)     .
\end{split} 
\end{equation}  
To obtain this we used instead of the radial derivative, the derivative operators $D_\ctor^\pm = \partial_\ctor \pm \frac{\beta\omega}{2}$, introduced in \cite{Jana:2020vyx}. They importantly satisfy the conjugation 
$e^{\beta\omega\ctor} D_\ctor^+ e^{-\beta\omega\ctor} = D_\ctor^-$, and also define the conjugate momentum in the non-orthonormal ingoing coordinate basis. 
The final step is realizing that the product of the ingoing propagator and the conjugate momentum $\pi$ is the boundary Green's function $K(k)$. This   
justifies our claims for the normalization of the bulk-bulk Green's function.

The bulk-bulk Green's function \eqref{eq:Gbulk} is normalizable on both ends of the grSK geometry. For fixed $\ctor'$, taking $\ctor \to 0$, picks up the $\GR{}(\ctor)\, \GL{}(\ctor') $ term, which vanishes by \eqref{eq:GLRbcs}. In the opposite limit $\ctor \to 1$ we pick up $\GL{}(\ctor) \, \GR{}(\ctor') $ which is again normalizable. Using the shorthand notation
\begin{equation}\label{eq:Thetanotation}
\begin{split}
\mathfrak{F}(\ctor_>) 
&= 
    \mathfrak{F}(\ctor)\, \Theta(\ctor-\ctor') +  \mathfrak{F}(\ctor')\, \Theta(\ctor'-\ctor) \,, \\
\mathfrak{F}(\ctor_<) 
&= 
    \mathfrak{F}(\ctor')\, \Theta(\ctor-\ctor') +  \mathfrak{F}(\ctor)\, \Theta(\ctor'-\ctor) \,, 
\end{split}
\end{equation}
we can express the bulk-bulk propagator in a compact form as 
\begin{equation}\label{eq:Gbulk2}
\Gbulk{a}(\ctor,\ctor';k)
= 
   \mathscr{N}(k)\,  e^{\beta\omega\,\ctor'}\, 
    \GL{}(\ctor_>, k)\, 
    \GR{}(\ctor_<', k) \,.
\end{equation}  

This completes the basic data necessary for computing correlation functions. Note that we only need to obtain the boundary-bulk propagator -- all the remaining Green's functions can be expressed in terms of it. This is indeed expected; for instance \cite{Herzog:2002pc} argued for an analogous structure in the thermofield double state. Their argument ought to apply to the Schwinger-Keldysh geometry. Likewise, we should highlight the fact that the  bulk-bulk Schwinger-Keldysh propagator for \SAdS{} black holes was written down in \cite{Arnold:2011hp} (see also \cite{Faulkner:2013bna}). Their result satisfies the inhomogeneous equation, but appears to be missing a radial Boltzmann factor, $e^{\beta\omega\,\ctor'}$. This factor, as we argue below, is crucial for ensuring that the boundary thermal correlators obey the Schwinger-Keldysh and KMS constraints.\footnote{
   The variation of parameters method does not a-priori fix the full functional dependence of the propagator on the source point ($X'$ in \eqref{eq:inhomeqn}). One way to proceed is to demand that the bulk-bulk propagator satisfies all the Schwinger-Keldysh and KMS constraints. Alternately, as we have done here, one can start with suitably normalized wavefunctions and fix the dependence using the Wronskian. \label{fn:fixexpbb}
} 

\subsection{Witten diagrams on the grSK geometry}
\label{sec:Wittendiag}

We now have all the ingredients necessary for computing a general $n$-point function with boundary Schwinger-Keldysh ordering. The recipe for computing contact diagrams has been described earlier in \cite{Jana:2020vyx}. We will supplement it here with the rules for computing the bulk exchange diagram. Additionally, we will argue that in neither case is there any contribution   localized on the horizon (modulo an assumption about vertex factors which we explain below).  

The general rule for computing diagrams is to start with the contour integral over the mock tortoise coordinate $\ctor$ and convert it to an integral over a single sheet of the spacetime outside the horizon. In doing so, we have to account for the radial Boltzmann factors, $e^{\beta \omega \zeta}$, arising from the outgoing and bulk-bulk propagator (including the piece originating from the Wronskian). Modulo these pieces, the rest of the integrand can be written in terms of $\Gin{}(\ctor)$, which is periodic \eqref{eq:Ginperiod}.  

Since the contour in \cref{fig:mockt} encircles the branch cut emanating from the horizon we typically are computing the integral of the discontinuity of a function evaluated on the two sheets. For instance, given a integrand $\mathfrak{L}(\ctor)$, which is regular on the horizon, it is easy to show that (nb: 
$d\ctor\, \sqrt{-g} = dr\, r^{d-1}$)
\begin{equation}
\oint \, d\ctor\, \sqrt{-g} \; \mathfrak{L}(\ctor) = \int_{r_+}^{r_c}\, dr\, r^{d-1} \bigg(\mathfrak{L}(\ctor(r)+1)  - \mathfrak{L}(\ctor(r))\bigg) \,.
\end{equation} 

Let us suppose for simplicity that we have an $n$-point self-interaction of a single field, say $\sen{}^n$. The bulk action contains a term of the form
\begin{equation}\label{eq:nvertex}
S_{(n)} \supset \oint d\ctor\, \int \,d^{d-1}x\, \sqrt{-g} \, e^{\chi_n(r)} \sen{}^n(\ctor,v,\vb{x})  \,. 
\end{equation} 
Expanding out the fields $\sen{}$ using the solution \eqref{eq:homsolFP} we can immediately write down an integral expression for influence functional $\mathcal{I}_{_{\text{F}\cdots\text{FP}\cdots \text{P}}}$, the term with say $p$ F sources, $\JF$ and $(n-p)$ P sources, $\JP$ in the retarded-advanced basis.
The integrand is a particular combination of the ingoing propagators and radial Boltzmann factors of the mock tortoise coordinate. 

Assembling the pieces we see that contour integral simply picks up a discontinuity, which owing to the periodicity of the ingoing Green's function \eqref{eq:Ginperiod} leads to an expression for the influence functional as a single-sheeted integral of the form \cite{Jana:2020vyx}\footnote{
  While we have described the situation for a contact interaction of single operator, the generalization to the case where the operators differ is  straightforward. 
}
\begin{equation}\label{eqn:Icontact}
\begin{split}
&
   \mathcal{I}_{_{\text{F}\cdots\text{FP}\cdots \text{P}}}(k_1,\cdots, k_n)
=
   \textrm{coeff}\bigg(\JF(k_1) \cdots \JF(k_p)\, \JP(k_{p+1}) \cdots \JP(k_n) \bigg) \\
&=
   \frac{(-1)^{p+1}}{p!\, (n-p)!} \, \left(1-e^{\beta\,\sum_{j={p+1}}^n\, \omega_j}\right) \int_{r_+}^{r_c}\, dr\, r^{d-1}\, \, e^{\chi_n(r)} \, \prod_{i=1}^p \, \Gin{}(\ctor,k_i) \prod_{j=p+1}^n \, e^{-\beta\omega_j \ctor}\, \Gin{}(\ctor,\wk_j) \,.
\end{split}
\end{equation} 

Note that the result depends on the analytic structure of the interaction vertex factor $e^{\chi_n(r)}$. In writing \eqref{eqn:Icontact} we assumed that this factor is regular the horizon.  This is, for instance, the case for the  minimally coupled fields, with non-derivative polynomial interactions. Vertex factors which have singularities at the horizon will lead to additional localized contributions. This, in fact, does occur.  For transverse fluctuations of a Nambu-Goto string probing the black hole such a vertex exists \cite{Chakrabarty:2019aeu}. One can argue that this extends to derivative interactions, which is the case for conserved currents. In addition, in those cases, it is also possible for there to be special kinematic loci where we find singular vertex functions, potentially along the integration contour. For simplicity, since we will assume the  vertex functions to be regular in our analysis, an assumption, which is reasonable to make for non-derivative polynomial interactions. The reader can find further commentary on this issue in \cref{sec:discuss} where we indicate some generalizations.

Let us next turn to exchange diagrams. The integrand in this case has not only the boundary-bulk propagators, but also the bulk-bulk propagator \eqref{eq:Gbulk}. If we expand out the latter, we can assemble the integrand as a product of ingoing boundary-bulk Green's functions (with some reversed frequencies) and radial Boltzmann factors of $e^{\beta \omega \ctor}$. We additionally have a product of contour-ordered step functions with various radial orderings, i.e., terms of the form  $\Theta(\ctor-\ctor')\, \Theta(\ctor'-\ctor'')$. We  decompose the calculation of the integral into a sum of terms, each with a single string of contour-order step function product. 

The number of these summands depends on the number of bulk-bulk propagators. A diagram with $\ell$ bulk-bulk propagators has $\ell+1$ distinguished radial interaction vertices which need to be ordered. Each bulk-bulk propagator has a binary ordering choice, so altogether we have $2^\ell$ terms each with a product of $(\ell-1)$ contour-ordered step functions arising when we expand out the integrand.

For instance, a single bulk exchange diagram, which has one bulk-bulk propagator has one contour-ordered step function, splits into a sum of two integrals, viz., 
\begin{equation}\label{eq:F12integrals}
\begin{split}
I_\text{1-ex} 
&= 
    \oint d\ctor \oint d\ctor' \bigg[ F_1(\ctor,\ctor')\, \Theta(\ctor-\ctor') + \  F_2(\ctor,\ctor')\, \Theta(\ctor'-\ctor) \bigg],
\end{split}
\end{equation}   
where the functions $F_1$ and $F_2$ are products of the propagators, radial Boltzmann factors, and vertex functions.  A two-exchange diagram will analogously result in four integrals as explained in \cref{sec:5pt}.

Each of these contour integrals needs to be evaluated by respecting the ordering specified. We use the contour-ordered step functions to select the relative ordering and convert the contour integral to a single copy integral.  For a given contour-ordered step function there are a-priori three sets of contributions: both the vertices on the L segment, both in the R segment, and one where there lie on opposite segments. In the latter case the two radial integrals run from the boundary to the horizon without constraint. However, when both are on the same sheet, one of the radial integrals is constrained by the other. To disentangle this, we adopt the following  convention:
\begin{itemize}[wide,left=0pt]
\item  $\Theta(\ctor-\ctor')$ is the contour ordered step function, with the flow dictated from the boundary SK contour, so out from $\Re(\ctor) =0$ towards $\Re(\ctor)=1$ encircling the cut.
\item $\theta(r-r')$ is the ordinary step function defined on a single sheet.  We express the radial integrals starting at the boundary and running to the horizon, to maintain consistency with the contour direction. Hence, $\theta(\ctor-\ctor') =1$ when $\ctor_c<\ctor'<\ctor$ and analogously for $\theta(\ctor'-\ctor)$. 
\end{itemize}

With this convention, we can convert the contour-ordering into  radial ordering in \eqref{eq:F12integrals}. We are furthermore free to use the step function identities
\begin{equation}\label{eq:stepidentity}
\Theta(\ctor-\ctor') + \Theta(\ctor'-\ctor) =1 \,, \qquad 
\theta(\ctor-\ctor') + \theta(\ctor'-\ctor) =1 \,, 
\end{equation} 
and convert all the constrained integrals into unconstrained ones. For instance, the reader can easily verify that the  two integrals in \eqref{eq:F12integrals} may be decomposed as{}
\begin{equation}\label{eq:4pt1exfull}
\begin{split}
I_\text{1-ex} 
&=
    \int_{\ctor_c}^{\ctor_+}\, d\ctor\,  \int_{\ctor_c}^{\ctor_+}\, d\ctor'\, \bigg(
        \mathfrak{F}_1 \, \theta(\ctor-\ctor') + \mathfrak{F}_2\, \theta(\ctor'-\ctor) \bigg)\,, \\
\mathfrak{F}_1 
&=
        F_1(\ctor,\ctor')- F_1(\ctor+1,\ctor') + F_2(\ctor+1,\ctor'+1) - F_2(\ctor,\ctor'+1) \,,\\
\mathfrak{F}_2
&=    
   F_1(\ctor+1,\ctor'+1)- F_1(\ctor+1,\ctor') + F_2(\ctor,\ctor') - F_2(\ctor,\ctor'+1) \,.
 \end{split}
\end{equation}  
The important fact to note is that the integrand is picking up appropriate discontinuities coming from the two sets of contour integrals folded down to a single-sheeted integral. We have organized the latter by the relative radial ordering of the vertices. This process can be iterated to compute any exchange diagram (see \cref{sec:exchangechecks} for further details). Additional features of exchange diagrams are also explained in \cite{Loganayagam:2022xyz}.

Schwinger-Keldysh and KMS conditions require that the influence functionals with purely retarded or advanced operator (i.e., purely P or F-type operators) should vanish. As noted in \cite{Jana:2020vyx}, for contact diagrams this trivially follows from the periodicity of the ingoing boundary-bulk propagator \eqref{eq:Ginperiod} (upon using momentum conservation for the case of all P sources). Therefore, the corresponding contribution from exchange diagrams also should vanish identically. This should follow again from the periodicity property and not the details of the boundary-bulk propagator. This can only happen if the combinations multiplying the radial step function combinations in the single-sheeted integrals are identically zero.  We have checked this to be the case for the four and five point functions with one and two exchanges, respectively, see \cref{sec:exchangechecks}. This, in fact, suffices for any integrand with one or two exchanges since the additional pieces are simply the boundary-bulk propagators (using momentum conservation to remove the exponential factor from the P sources), which are themselves periodic. It should be possible to generalize this argument to prove that the result holds for any number of exchanges, and also for bulk loops, which we leave for the future. 
  
One can also give a unified presentation of the non-vanishing Schwinger-Keldysh correlators. For instance, the single-exchange four-point functions involving four different external operators $\{\mathcal{O}_a,\mathcal{O}_b,\mathcal{O}_c, \mathcal{O}_d\}$, and the operator $\mathcal{O}_{e}$ being exchanged can be obtained from the integral
\begin{equation}\label{eq:4ptmaster}
\begin{split}
\IM{abcde}{ij}{k_1,k_2,k_3,k_4}{k_5,k_6,\omega_7,\omega_8} 
&= 
   \int_{\ctor_c}^{\ctor_+}\, d\ctor\, \sqrt{-g}\, e^{\chi_i(\ctor)}\, e^{\omega_7 \ctor} \, \Gin{a}(\ctor,k_1)\, \Gin{b}(\ctor,k_2)\, \Gin{e}(\ctor,k_5) \\
&\quad 
   \times \int_{\ctor_c}^{\ctor}\,d\ctor' \sqrt{-g}\, e^{\chi_j(\ctor')} \, e^{\omega_8 \ctor'} \, \Gin{c}(\ctor',k_3)\, \Gin{d}(\ctor',k_4)\, \Gin{e}(\ctor',k_6)\,.
\end{split}
\end{equation}
The integrand is expressed solely in terms of the ingoing boundary-bulk propagator using \eqref{eq:Gout}, \eqref{eq:Grev},  and \eqref{eq:Gbulk}. The two factors of $\Gin{e}$ originate from the bulk-bulk propagator and are kept distinct by the use of the different frequency labels. The superscript species label indicates the dependence on the index $\ann$ and the dimension $\Delta$ of the external and internal operators, while the explicit $e^{\chi(\ctor)}$ factors allow for potential asymmetry of the vertices.  The radial Boltzmann weight factors $e^{\omega_7\,\ctor}$ and  $e^{\omega_8\,\ctor'}$ originate from either outgoing propagators or the Wronskian factor of the bulk-bulk propagator. In physical examples, $k_5,k_6,\omega_7,\omega_8$ are functions of the external momenta, but it is helpful to distinguish them for clarity. 

We give here the final expressions for the four-point functions in terms of the master integral defined in \eqref{eq:4ptmaster}. To keep the expressions compact we will write them in terms of the exchanged momenta  $k = k_3+k_4 = -(k_1  + k_2)$ whose frequency is $\omega$. 

Assuming the external operators to be all distinct, there are a-priori three correlators we should consider three orderings: FFFP, FFPP, and FPFP. All others can be obtained by exchanges $F\leftrightarrow P$. First we note that the FFFP influence function takes the form
\small
\begin{equation}\label{eq:FFFP}
\begin{split}
&\mathcal{I}^{abcd}_{_\text{FFFP}}(k_1,k_2,k_3,k_4) 
=
   \mathscr{N}(k)\,\frac{\nB(\omega)+1}{\nB(\omega_4)} 
   \bigg[
    \IM{abcde}{ij}{k_1,k_2,k_3,\wk_4}{\wk,\wk,-\omega,-\omega_4}  -\IM{abcde}{ij}{k_1,k_2,k_3,\wk_4}{\wk,k,-\omega,\omega_3}\\
&
\hspace{5.5cm} 
   + \IM{cdabe}{ji}{k_3,\wk_4,k_1,k_2}{\wk,\wk,-\omega_4,-\omega}
   -\IM{cdabe}{ji}{k_3,\wk_4,k_1,k_2}{\wk,k,-\omega_4,0}
   \bigg] \,.
\end{split}
\end{equation}
The FFPP correlator is similarly simple, and can be shown to be 
\begin{equation}\label{eq:FFPP}
\begin{split}
\mathcal{I}^{abcd}_{_\text{FFPP}}(k_1,k_2,k_3,k_4) 
&= 
    \mathscr{N}(k)\,e^{\beta\omega}
 \bigg[
   \IM{abcde}{ij}{k_1,k_2,\wk_3,\wk_4}{\wk,k,-\omega,0} -\;\IM{abcde}{ij}{k_1,k_2,\wk_3,\wk_4}{\wk,\wk,-\omega,-\omega} \\ 
& \qquad \qquad \qquad  
   + \IM{cdabe}{ji}{\wk_3,\wk_4,k_1,k_2}{\wk,k,-\omega,0}
      - \IM{cdabe}{ji}{\wk_3,\wk_4,k_1,k_2}{\wk,\wk,-\omega,-\omega}  \bigg] \,.
\end{split}
\end{equation}
Finally, the FPFP correlator, which happens to be the most involved, turns out to be 
\begin{equation}\label{eq:FPFP}
\begin{split}
\mathcal{I}^{abcd}_{_\text{FPFP}}(k_1,k_2,k_3,k_4) &= 
   \mathscr{N}(k)\,(\nB(\omega)+1)^2 \left[ \frac{ \mathfrak{A}_1}{\nB(\omega_2)\, (\nB(\omega_3) +1)}  - e^{\beta\omega_2} \, \frac{ \mathfrak{A}_2}{\nB(\omega_1)\, \nB(\omega_4)}  \right]  \\
\mathfrak{A}_1
&= 
      \IM{abcde}{ij}{k_1,\wk_2,k_3,\wk_4}{k,k,-\omega_2,\omega_3} 
      - \IM{abcde}{ij}{k_1,\wk_2,k_3,\wk_4}{k,\wk,-\omega_2,-\omega_4} \\
&\qquad \qquad 
      +  \IM{cdabe}{ji}{k_3,\wk_4,k_1,\wk_2}{k,k,\omega_3,-\omega_2} 
      - \IM{cdabe}{ji}{k_3,\wk_4,k_1,\wk_2}{k,\wk,\omega_3,\omega_1}    \\
\mathfrak{A}_2
&= 
    \IM{abcde}{ij}{k_1,\wk_2,k_3,\wk_4}{\wk,k,\omega_1,\omega_3} - \IM{abcde}{ij}{k_1,\wk_2,k_3,\wk_4}{\wk,\wk,\omega_1,-\omega_4}   \\
&\qquad \qquad
    +\IM{cdabe}{ij}{k_3,\wk_4,k_1,\wk_2}{\wk,k,-\omega_4,-\omega_2} 
      - \IM{cdabe}{ij}{k_3,\wk_4,k_1,\wk_2}{\wk,\wk,-\omega_4,\omega_1}  \,.
\end{split}
\end{equation}
\normalsize

In writing the expressions we have repeatedly used momentum conservation for simplification. By assuming that the external operators are all distinct, and allowing the vertices to be non-identical, we have avoided having to sum over different channels. In the case of identical external operators we should add contributions form different channels as appropriate.

The arguments $\omega_7$ and $\omega_8$ of the master integral \eqref{eq:4ptmaster} which enter into the expressions for the four-point correlators \eqref{eq:FFFP}-\eqref{eq:FPFP} are not generic, but are constrained to satisfy
\begin{equation}\label{eq:om78constrained}
\omega_7 = \frac{1}{2}\left(\omega_1+ \omega_2 + \omega_5\right) \,, 
\qquad
\omega_8 = \frac{1}{2}\left(\omega_3+ \omega_4 + \omega_6\right) \,. 
\end{equation} 
This fact will be helpful when we attempt an explicit evaluation of the integral for the two-dimensional model we introduce.

\subsection{Analytic structure of the correlators}
\label{sec:analcorr}

We are now in a position to understand the general features of thermal real-time correlators computed using holography. We delineate the analytic structure of the higher-point functions computed using contact and exchange diagrams. We spell out the assumptions we are making to deduce these results as we go along. 

Consider the ingoing boundary-bulk propagator $\Gin{}$, which obeys \eqref{eq:Gin}. In particular, the source has been normalized to unity. Black hole quasinormal modes, on the other hand, are defined to be normalizable modes that are ingoing at the horizon. The ingoing propagator therefore ought to have poles at the quasinormal frequencies.  Based on this intuition it will be useful to write down a factorized form for the boundary-bulk propagator
\begin{equation}\label{eq:GKfactor}
\Gin{}(\ctor,k) = \Kin{}(k)\, \Gint{}(\ctor,k)\,.
\end{equation} 

The function $\Kin{}(k)$ is the retarded thermal two-point function on the boundary.  $\Kin{}(k)$ is meromorphic, its poles are the quasinormal modes, which for a sensible thermal system, reside in the lower half of the complex frequency plane. The function $\Gint{}(\ctor,k)$ is generically regular. It has zeros in its non-normalizable part to compensate for the $\Kin{}$ we factored out. This is because we have chosen to normalize the source to unity on the boundary. Its normalizable part is clearly analytic, as it actually corresponds to the boundary correlator.

When the boundary-bulk propagator is used to compute the boundary two-point function, one ends up evaluating the difference of the product of the field and its conjugate momentum, $\sen{} \pi$, at the two asymptotic boundaries of the grSK geometry. Using the asymptotic behaviour \eqref{eq:Gin} and \eqref{eq:conjmom} one then derives the retarded boundary correlator $\Kin{}(k)$. This typically takes the form of a rational function. While this is guaranteed by virtue of meromorphicity, in examples one finds that the structure to be more specific. The numerator is a function $\gfn{}(k)$ which depends on the conformal dimension, and the denominator ends up being the same function, but now evaluated as a function of the shadow dimension.  The Schwinger-Keldysh structure is completely captured by the fact that only the  $\JF\, J_d$ combination of the source term is present in the product, cf., \eqref{eq:s2PF}. 
For minimally coupled scalar fields, expressions for $\Kin{}(k)$ have been known in the \AdS{3}/CFT$_2$ context for a long while. They were originally discussed in \cite{Birmingham:2001pj} by analytically continuing CFT results, but also later obtained using real-time methods in \cite{Son:2002sd}.\footnote{
   They were rederived using the grSK geometry in \cite{Jana:2020vyx} where the aforementioned structure can be readily verified. In a certain sense, the grSK geometry proposed in \cite{Glorioso:2018mmw} gives us a much cleaner way to arrive at the result (a fact which was already appreciated in \cite{vanRees:2009rw}).}
Likewise, the result for $\Kin{}(k)$ is now also known in four dimensions \cite{Dodelson:2022yvn} using connection between the minimally coupled scalar wave equation in \SAdS{5} and differential equations satisfied by supersymmetric instanton partition functions. 

This data is sufficient to deduce the analytic structure of the higher-point correlation functions. As discussed in \cref{sec:Wittendiag}, we will assume here that the vertex functions $e^{\chi_\lambda}$ are non-singular along the ray $\Re(r) \in (r_+,\infty)$ running from the horizon to the boundary.

\paragraph{Contact diagrams:}  Consider first bulk contact interactions; we can use \eqref{eqn:Icontact} to deduce the analytic structure of the influence functionals. The integrand is a product of ingoing and outgoing propagators. Using the decomposition \eqref{eq:GKfactor}, ingoing propagators have quasinormal poles and outgoing propagators have anti-quasinormal poles (due to frequency reversal). This implies we can rewrite \eqref{eqn:Icontact} as 
\begin{equation}\label{eqn:IcontactGK}
\begin{split}
\mathcal{I}_{_{\text{F}\cdots\text{FP}\cdots \text{P}}}(k_1,\cdots, k_n)
&=
     \left[\prod_{i=1}^p \, \Kin{}(k_i) \, \prod_{j=p+1}^n \, \Kin{}(\wk_j) \right] \, \widetilde{\mathcal{I}}_{_{\text{F}\cdots\text{FP}\cdots \text{P}}}(k_1,\cdots, k_n) \,, \\
\widetilde{\mathcal{I}}_{_{\text{F}\cdots\text{FP}\cdots \text{P}}}(k_1,\cdots, k_n) 
&= 
   \frac{(-1)^{p+1}}{p!\, (n-p)!}  \left(1-e^{\beta\,\sum_{j={p+1}}^n\, \omega_j}\right) \\
&\quad  \times    
    \int_{r_+}^{r_c}\, dr\, r^{d-1}\, \, e^{-\chi_n(r)} \, \prod_{i=1}^p \, \Gint{}(\ctor,k_i) \prod_{j=p+1}^n \, e^{-\beta\omega_j \ctor}\, \Gint{}(\ctor,\wk_j) \,.
\end{split}
\end{equation} 
We have factored out all the non-analytic pieces in the first line. The integrand of $\widetilde{\mathcal{I}}$ in the second line is a product of analytic functions $\Gint{}(k)$. As long as the radial integral converges in the limit $r_c\to \infty$, and at the lower limit $r_+$ (the latter is at times more constraining),  we expect no additional non-trivial singularities. By this we mean that we should not encounter singularities that depend on the dynamical data, viz., momenta and dimensions of the external operators. 

It is however possible that owing to the presence of the radial Boltzmann weights we end up with a function that is cognizant of the statistical factor $\nB$. First, we note that 
\begin{equation}\label{eq:nbfactor}
\nB(\omega) \propto \frac{1}{\sinh\left(\frac{\beta\omega}{2}\right)} 
\propto \Gamma\left(1+i\,\frac{\beta\omega}{2\pi}\right) \, \Gamma\left(1-i\,\frac{\beta\omega}{2\pi}\right)
\end{equation} 
Given a particular sequence of F or P operators,  by causality the correlator for the corresponding $\omega_i$ should be analytic in the upper or lower half-plane, respectively. In \eqref{eqn:IcontactGK} the Boltzmann factors involve a sum of advanced frequencies, and thus any contribution proportional to them should be analytic in the lower-half  $\omega_j$ planes, with $j=p+1, \cdots, n$ plane. Should the radial integral produce a non-analytic term, it has to, for consistency respect this. We indeed find that the radial integrals generically produces a factor of $\Gamma\left(1 + \frac{i\beta}{2\pi}\,  \sum_{j={p+1}}^n\, \omega_j\right)$. This has the correct analytic structure -- its poles are at Matsubara frequencies of $\sum \omega_j$ in the upper half-plane. 

However, the presence of such Matsubara poles in the external operator kinematics is forbidden by the fact that the FP-basis factors out the KMS constraints. Indeed, this was the primary motivation for the introduction of the FP basis in \cite{Chaudhuri:2018ihk}. They in particular, argued that the Schwinger-Keldysh correlators can be obtained from a suitable $n$-point spectral function, dressed up with Boltzmann factors to obtain the desired correlator specified by F and P labels.

In contact diagram, these Gamma factors with Matsubara poles are rendered harmless by the overall statistical factor out front in \eqref{eqn:IcontactGK}, whose zeros nullify the poles.  This, in fact, serves as a nice cross-check for the vanishing of the all F or all P correlator. In summary, the contact diagrams indicate that the $n$-point correlator only has quasinormal poles corresponding to the retarded (F) operators, and anti-quasinormal poles corresponding to the advanced (P) operators.

\paragraph{Exchange diagrams:} These ought to behave similarly, but here we need to account for the analog of unitarity cuts arising from the poles of the bulk-bulk propagator.
Let us first use the factorized form to rewrite the bulk-bulk propagator:
\begin{equation}\label{eq:Gbulkexp}
\begin{split}
\Gbulk{}(\ctor,\ctor',k) 
&=   -\nB\, \left(\nB+1\right) \mathscr{N}(k)\;  e^{\beta\omega\,\ctor'} 
   \bigg[\Theta(\ctor - \ctor') \, 
   \left(\Kin{}(k)\, \Gint{}(\ctor',k) -  \Kin{}(\wk)\, \Goutt{}(\ctor',k)\right) \\
& \qquad \qquad \qquad \qquad \times 
   \left(\Kin{}(k)\, \Gint{}(\ctor,k) -e^{\beta\omega}\, \Kin{}(\wk)\, \Goutt{}(\ctor,k)\right)
   +\ctor \leftrightarrow \ctor'  \bigg] \,.
\end{split}
\end{equation} 

Then note that the normalization factor $\mathscr{N}(k)$, which is given by \eqref{eq:Nkgeneric} has a factor of $\Kin{}(k)-\Kin{}(\wk)$ in the denominator. It also has a set of Matsubara zeros owing to the Boltzmann factor $\nB+1$.
This implies that $\mathscr{N}(k)$ has both quasinormal and anti-quasinormal poles -- this follows since $\Gin{}$ and $\Gout{}$ furnish a basis of solutions.\footnote{
   Equivalently, we could have used the fact that $\GL{}$ and $\GR{}$ have both quasinormal and anti-quasinormal poles, which is necessary since neither set by itself is a complete basis of mode solutions \cite{Warnick:2013hba}.  
} 
For purposes of ascertaining the analytic structure, we can replace 
$(\Kin{}(k)-\Kin{}(\wk))^{-1} $ by $(\Kin{}(k)\, \Kin{}(\wk))^{-1}$.
With this understanding, it then follows that one of the step-function terms in $\Gbulk{}$ takes the form 
\begin{equation}\label{eq:Gbulkpoles}
\begin{split}
\Gbulk{}(\ctor,\ctor',k) \bigg|_{\ctor > \ctor'}
& \propto \nB \left(
   \Kin{}(k)\, \Gint{}(\ctor,k)\, \Gint{}(\ctor',k)
   +  \Kin{}(\wk)\, \Goutt{}(\ctor,k)\, \Goutt{}(\ctor',k) \right.\\
& \left.\qquad
   - e^{\beta\,\omega}\, \Gint{}(\ctor',k)\, \Goutt{}(\ctor,k)
    -  \Gout{}(\ctor,k)\, \Gint{}(\ctor',k) \right) \,.
\end{split}
\end{equation} 
Therefore, the only terms from the bulk-bulk propagator which have singularities are when both factors are ingoing or outgoing. The former has quasinormal poles, while the latter has anti-quasinormal poles.

It is easy to see what this property of the bulk-bulk propagator implies for an arbitrary tree level exchange. To do so, let us generalize the master integral  by  introducing a primitive integrand for cubic bulk vertices\footnote{
   The generalization to higher order bulk interactions is straightforward; we simply upgrade the primitive to include as many factors of the boundary-bulk propagator as the valency of the vertex.
}
\begin{equation}\label{eq:primitive}
\begin{split}
\mathfrak{B}_i^{abc}(k_1,k_2,k_3;\ctor)  
&= 
      \Kin{a}(k_1)\, \Kin{b}(k_2)\, \Kin{c}(k_3) \; \widetilde{\mathfrak{B}}_i^{abc}(k_1,k_2,k_3;\ctor) \\
\widetilde{\mathfrak{B}}_i^{abc}(k_1,k_2,k_3;\ctor)  \,,
&=
       \sqrt{-g}\, e^{\chi_i(\ctor)} \, \Gint{a}(\ctor,k_1)\, \Gint{b}(\ctor,k_2)\, \Gint{c}(\ctor,k_3) \,.
\end{split}
\end{equation} 
The function $\widetilde{\mathfrak{B}}$ is now analytic, with the poles factored out. A general tree level exchange requires the evaluation of a recursive integral of the form  
\begin{equation}\label{eq:recurseint}
\begin{split}
& \int_{\ctor_c}^{\ctor_+} \, d\ctor_1 \, 
   \widetilde{\mathfrak{B}}_{i_1}^{a_1b_1c_1}(k^{(1)}_1,k^{(1)}_2,k^{(1)}_3;\ctor_1) \; 
   \int_{\ctor_c}^{\ctor_1} \, d\ctor_2 \, 
      \widetilde{\mathfrak{B}}_{i_2}^{a_2b_2c_2}(k^{(2)}_1,k^{(2)}_2,k^{(2)}_3;\ctor_2) \cdots\cdots \\ 
& \qquad \qquad 
   \times \cdots \cdots 
   \int_{\ctor_c}^{\ctor_{\ell-1}} \, d\ctor_\ell \, 
      \widetilde{\mathfrak{B}}_{i_\ell}^{a_\ell b_\ell c_\ell}(k^{(\ell)}_1,k^{(\ell)}_2,k^{(\ell)}_3;\ctor_\ell)\,.
\end{split}
\end{equation} 
This integral, should it converge, is analytic, even with some external or internal frequencies reversed. All the factors of $\Kin{}(k)$ which account for the quasinormal poles, neatly factor out. We then learn that a tree-level exchange has
\begin{itemize}[wide,left=0pt]
\item  Quasinormal (or anti-quasinormal) poles from the factors $\Kin{}(k)$ multiplying the external boundary-bulk propagators.
\item A set of quasinormal and anti-quasinormal poles for each of the internal bulk-bulk propagators, arising from two of the terms in \eqref{eq:Gbulkpoles}. This occurs when successive factors of $\widetilde{\mathfrak{B}}$ in the integral \eqref{eq:primitive} have their third momentum argument coincide. The poles are in the lower half-plane when both the propagators from the bulk-bulk Green's function are ingoing, and in the upper half-plane when they are outgoing.  
\item Potentially additional Matsubara poles in the exchanged frequencies. These are not forbidden by the spectral analysis of \cite{Chaudhuri:2018ymp}, but have to respect the causality properties of the correlator. If the exchanged frequency arises from all F (or all P) operators, then we could have Matsubara poles in the lower (or upper) half-planes. If both F and P type operators are involved, then the poles can occur in the entire frequency plane.
\end{itemize}

It is easy to check then that the four-point influence functionals given earlier have the following analytic structure:
\begin{itemize}[wide,left=0pt]
\item  $\mathcal{I}^{abcd}_{_\text{FFFP}}(k_1,k_2,k_3,k_4)$: quasinormal poles in the external  $\omega_1,\omega_2,\omega_3$ and anti-quasinormal poles in $\omega_4$. Additionally, there are-quasinormal and Matsubara poles in $\omega_1+\omega_2$ (equivalently, anti-quasinormal modes in $\omega_3 + \omega_4$) from the first and third master integrals in \eqref{eq:FFFP}. 
\item  $\mathcal{I}^{abcd}_{_\text{FFPP}}(k_1,k_2,k_3,k_4)$: quasinormal poles in $\omega_1,\omega_2$ and anti-quasinormal poles in $\omega_3,\omega_4$. Additionally, there are quasinormal poles in $\omega_1+\omega_2$ from the second and fourth master integral terms of \eqref{eq:FFPP}.
\item  $\mathcal{I}^{abcd}_{_\text{FPFP}}(k_1,k_2,k_3,k_4)$: quasinormal poles in $\omega_1,\omega_3$ and anti-quasinormal poles in $\omega_2,\omega_4$. Additionally, there are poles in $\omega_3+\omega_4$ or $\omega_1+\omega_2$ from the four of the eight master integrals in \eqref{eq:FPFP}. In $\mathfrak{A}_1$ the first and third terms give quasinormal poles for $\omega_3+\omega_4$ while in $\mathfrak{A}_2$ the second and fourth give quasinormal poles in $\omega_1+\omega_2$. There are also Matsubara poles in $\omega_3+\omega_4$ in the entire complex plane. 
\end{itemize}
We shall verify these statements explicitly below when we compute the correlators in our two-dimensional example. The discussion above does not rely on specifics of the model, and only rests on the structure of the Green's functions and bulk interaction vertices.

\section{A low dimensional toy model}
\label{sec:BTZtoy}

The discussion above has been quite general and serves to highlight the fact that the computation of real-time thermal correlation functions from holography is in principle on as firm a footing as computing vacuum correlators. 
It would however be useful to have an example where we can compute the correlation functions explicitly. In dimensions $d>2$, there are no analytic expressions available for the essential ingredient, the boundary-bulk propagator (for asymptotically \AdS{d+1} black holes). However, the BTZ black hole in $d=2$ provides a setting which is analytically tractable. This was already exploited in \cite{Jana:2020vyx} to compute the two and three point influence functionals for a minimally coupled massive scalar field. Generalization to the designer fields is straightforward, and as we shall see below, allows us to even mimic the behaviour of non-Markovian dynamics observed for conserved currents in the higher dimensional examples. In the rest of the paper we will use this simple toy model to illustrate the general principles.

\subsection{The set-up}
\label{sec:setup}

We aim to study a simple model of designer scalar fields in the BTZ geometry. We will work with the action \eqref{eq:Sdesign} for the fields $\sen{a}$, characterized by indices $\ann_a$ and dimension $\Delta_a$, respectively. As described in \cref{sec:sprops} we treat the $\Delta_a$ as a proxy for the masses of the bulk fields. While they do not necessarily indicate the conformal dimension of a boundary operator in the dual 2d CFT, we will find it convenient  to refer to them as such.

The line element for the BTZ black hole in ingoing coordinates is given by
\begin{equation}\label{eq:btzmet}
\begin{split}
ds^2 
&= 
   - r^2 \left(1-\frac{r_+^2}{r^2}\right) \, dv^2 + 2 \, dv \, dr + r^2\, dx^2 \\
&=
   \frac{1}{z^2} \bigg[
   -(1-z^2)\, dv^2 - 2\, dv \, dz + dx^2
   \bigg] \,.
\end{split}
\end{equation} 
In the second line we have introduced the inverse radial coordinate $z \equiv \frac{r_+}{r}$. The solution has a Hawking temperature $T = \frac{r_+}{2\pi}$. The mock tortoise coordinate can be determined by integrating \eqref{eq:ctordef}, and we find 
\begin{equation}\label{eq:ctorbtz}
\frac{r}{r_+} = \frac{1}{z} =  \tan\left(\frac{\pi}{2} - \pi \, \ctor\right) 
\end{equation} 

The dynamics of a designer field $\sen{}$ with index $\ann$ and dimension $\Delta$ is determined by the equation of motion 
\begin{equation}\label{eq:btzMeqn}
-\frac{z^3}{(1-z^2)\, \, e^{\chi(z)}} \, \Dz_+ \left(\frac{e^{\chi(z)}}{z}\, \Dz_+ \sen{\ann}(z) \right) + z^2 \, \left(\bqt^2 - \frac{\bwt^2}{1-z^2}\right)\, \sen{\ann}(z)  + m^2 \, \sen{\ann}(z) = 0 
\end{equation} 
The auxiliary dilaton is taken to be a power law $e^\chi = z^{1-\ann}$. 
We have written the equation in terms of time-reversal covariant derivative $\Dz_+$, cf., \cite{Ghosh:2020lel}, and introduced the dimensionless frequency and momenta $\bwt$ and $\bqt$, respectively. These are defined to be
\begin{equation}\label{eq:Dpwq}
\Dz_+ = - (1-z^2)\, \dv{z} - i\, \bwt \,, \qquad \bwt = \frac{\omega}{r_+} \,, \qquad \bqt = \frac{\abs{\vb{k}}}{r_+} \,.
\end{equation} 

The solution to the wave equation \eqref{eq:btzMeqn} can be readily obtained in terms of hypergeometric functions. Imposing regularity at the horizon and a unit normalized source on the boundary for a field $\sen{a}$ obeying Dirichlet boundary conditions, we find the ingoing boundary-bulk Green's function to be\footnote{
   In writing down the solution we are assuming that $\Delta -\tilde{\Delta} \notin \mathbb{Z}$. When this difference is integral the solutions include a logarithmic branch. The regularized hypergeometric function are boldfaced \cite[15.1.2,16.2.5]{NIST:DLMF}; $\pFqReg{2}{1}{\bpt_+ + \frac{\Delta}{2}, \bpt_-+\frac{\Delta}{2}}{ 1-i\,\bwt}{1-z^2}$ appearing in \eqref{eq:Ginbtz} is regular in the domain $z\in [0,1)$.
} 
\begin{equation}\label{eq:Ginbtz}
\Gin{\ann,\Delta}(z, k) = 
   \frac{\Gamma(\bpt_+ + \frac{\Delta}{2})\, \Gamma(\bpt_-+ \frac{\Delta}{2})}{\Gamma\left(\Delta - \frac{1+\ann}{2}\right)} \,
    z^\Delta \, (1+z)^{-i\,\bwt} \, \pFqReg{2}{1}{\bpt_+ + \frac{\Delta}{2},\bpt_- + \frac{\Delta}{2}}{1-i\,\bwt}{ 1-z^2}\,.
\end{equation} 
This solution can be uplifted to the full grSK geometry with the aid of \eqref{eq:ctorbtz}.

In presenting the solution, we introduced the combination $\bpt_\pm$, which are defined as 
\begin{equation}\label{eq:ppmdef}
\bpt_\pm = \frac{i}{2}\, \left(-\bwt \pm \sqrt{\bqt^2 - \frac{(\ann-1)^2}{4}}\right)   - \frac{\ann-1}{4} \,.
\end{equation} 
For a minimally coupled scalar field, which has $\ann =1$ (in $d=2$) these combinations reduce to the light-cone momenta. For non-minimally coupled designer scalars, however, we notice that the dependence on the spatial momentum is modified to a surdic form. This observation will be of importance in modeling non-Markovian dynamics with our model. 
For later reference let us also record the symbol for the frequency reversed version of $\bpt_\pm$, which we denote with a bar decoration (as for the 3-momentum $k$) 
\begin{equation}\label{eq:palt}
\bpts_\pm = \frac{i}{2}\, \left(\bwt \pm \sqrt{\bqt^2 - \frac{(\ann-1)^2}{4}}\right)  - \frac{\ann-1}{4} \,.
\end{equation} 

The final piece of data we need is the bulk-bulk propagator which we can express in terms of the ingoing boundary-bulk propagator following \eqref{eq:Gbulk}. We find 
\begin{equation}\label{eq:Gbulkbtz}
\Gbulk{\ann, \Delta}(\ctor,\ctor';k) = 
   \mathscr{N}_{\ann,\Delta}(k) \, e^{2\pi\,\bwt\,\ctor' } \,
    \GL{\ann,\Delta} (\ctor_>,k)\, \GR{\ann,\Delta} (\ctor_<,k) \,,
\end{equation} 
with the normalization determined  to be
\begin{equation}\label{eq:Nfix}
\mathscr{N}_{\ann,\Delta} (k) = -\pi i\, \frac{e^{-\pi \,\bwt} \, \Gamma\left(\Delta - \frac{1+\ann}{2}\right)^2}{\Gamma(\bpt_+ +\frac{\Delta}{2}) \, \Gamma(\bpt_-+\frac{\Delta}{2}) \, \Gamma(\bpts_++\frac{\Delta}{2})\, \Gamma(\bpts_-+\frac{\Delta}{2})} \,.
\end{equation} 
We will confirm later that this agrees with the form quoted in \eqref{eq:Nkgeneric}.

The results above are derived for the case where the field $\sen{\ann}$ is quantized with Dirichlet boundary condition. If we were to instead quantize it using Neumann boundary condition (depending on the relative values of $\ann$ and $\Delta$ as described in \cref{sec:sprops}), the only change is that we replace the conformal dimension $\Delta$ with the appropriate generalization of the shadow dimension 
\begin{equation}\label{eq:shadowD}
\tilde{\Delta} = 1+\ann - \Delta \,.
\end{equation} 
In particular, the ingoing boundary-bulk Green's function with Neumann boundary condition is obtained from \eqref{eq:Ginbtz} by swapping $\Delta$ and $\tilde{\Delta}$. We therefore for the most part will focus on the Dirichlet choice, and use the replacement rule $\Delta \leftrightarrow 1+\ann - \Delta$ to infer the answer for the Neumann case. 

The answers obtained here for the dilaton modulated scalar in the BTZ geometry, curiously enough, have also been realized earlier. In  \cite{Davison:2014lua} a three-dimensional theory with broken spatial translations was studied. The authors studied an Einstein-scalar system, with two scalars having profiles that preserve the spatial symmetries in the bulk, while having non-vanishing, symmetry breaking, sources along the boundary. The dual is planar black hole with non-degenerate horizon. Unlike the \SAdS{4} geometry, the presence of non-vanishing scalar profiles, causes the metric to fall-off more slowly, with the emblackening function being of the BTZ form.\footnote{
   The bulk metric has $f(r) = 1-\frac{r_+^2}{r^2}$ in $d=3$. Therefore, technically, the geometry is not asymptotically locally \AdS{4} in the standard sense, since the fall-offs are too slow. However, they can be viewed as such, since accounting for scalar counterterms the boundary observables are finite. 
} 
If we study probe fields in this geometry that are insensitive to the broken translations (i.e., those that do not couple to the scalar fields sourcing the geometry) then the probe equations will reduce to \eqref{eq:btzMeqn} for some $\ann$. \cite{Davison:2014lua} analyzed probe Maxwell dynamics in this spacetime, which indeed behaves as we have seen above -- the wavefunctions are hypergeometric, and the momentum dependence is of the surdic form in $\bpt_\pm$. The paper also examined certain components of the stress tensor correlators, but here we are mildly confused about the harmonic plane wave decomposition in a system with broken translations. This model later used to illustrate the convergence of the hydrodynamic expansion of $U(1)$ current correlators in \cite{Grozdanov:2019uhi}.

\subsection{Gaussian influence functionals}
\label{sec:gaussian}

Armed with the boundary-bulk propagators we can evaluate the on-shell action at  quadratic order to obtain the real-time two-point functions of the dual operator $\mathcal{O}$. 

First, consider the case where we choose Dirichlet boundary conditions for the field $\sen{}$. We find that the on-shell action evaluates to a boundary contribution 
\begin{equation}\label{eq:SoSeval}
S = \bigg[ \frac{1}{2} \int \, d^2x\,\, \pi\, \sen{} + S_\text{ct} \bigg]
_{\ctor=0}^{\ctor=1} \,.
\end{equation} 

The conjugate momentum $\pi$ has an asymptotic behavior
\begin{equation}
\lim_{z\to 0} \, \pi_{\text{ren}} = z^{\Delta -1-\ann} \, \Kin{\ann,\Delta}(k)  + \cdots \,.
\end{equation} 
The kernel $\Kin{\ann,\Delta}(k)$, which is the boundary retarded correlator, is given as the ratio of Gamma functions\footnote{
    The overall normalization factor in $\Kin{\ann,\Delta}$ should be computed with care. Naively, one expects the conjugate momentum to have a factor of $\Delta$ from the radial derivative. However, this gets converted to the quoted factor when a proper counterterm renormalization is carried out. 
}
\begin{equation}\label{eq:Kindef}
\begin{split}
\Kin{\ann,\Delta}(k) 
&=
    2 \left(\Delta - \frac{\ann+1}{2}\right)\, \frac{\gfn{\ann}(k, \Delta)}{\gfn{\ann}(k,\tilde{\Delta})} \,,\\
\gfn{\ann}(k,\Delta) 
&\equiv 
   \Gamma\left(\bpt_++\frac{\Delta}{2}\right) \, \Gamma\left(\bpt_-+\frac{\Delta}{2}\right) \, \Gamma\left(\frac{1+\ann}{2} - \Delta\right). 
\end{split}
\end{equation} 
Indeed, the generating function for the boundary correlators reads:
\begin{equation}\label{eq:s2PF}
S_{\ann,\Delta}^{(2)} [\JF,\JP]
= -\frac{1}{2} \, \int_k \, \frac{1}{\nB} \, \JP(k)\, \Kin{\ann,\Delta}(k) \, \JF(-k) \,.
\end{equation} 
This can be expressed in a more familiar form by switching to the average difference basis of operators 
\begin{equation}\label{eq:avdif}
J_a = \frac{1}{2}\left(J_\skR + J_\skL\right) \,, \qquad J_d = J_\skR - J_\skL \,.
\end{equation} 
The quadratic effective action then takes the form
\begin{equation}\label{eq:S2ad}
S_{\ann,\Delta}^{(2)} [J_a,J_d]
= -\frac{1}{2} \, \int_k \, J_d(-k)\, \Kin{\ann,\Delta}(k) \,  
   \left[J_a(k) + \left(\nB(\omega)+\frac{1}{2}\right)\, J_d(k) \right] ,
 \end{equation} 
making it transparent that the kernel indeed gives the retarded two-point function (it is the coefficient of the source term $J_a\, J_d$). 

Now that we have the boundary retarded correlator we can confirm that the normalization of the bulk-bulk propagator quoted in \eqref{eq:Nfix} has the correct analytic structure. We find
\begin{equation}\label{eq:Nkrelation2d}
\Kin{\ann,\Delta}(k) - \Kin{\ann,\Delta}(\wk) = \frac{2i}{\pi}\, \sinh(\pi\,\bwt)\, 
\frac{\Gamma(\bpt_+ +\frac{\Delta}{2}) \, \Gamma(\bpt_-+\frac{\Delta}{2}) \, \Gamma(\bpts_++\frac{\Delta}{2})\, \Gamma(\bpts_-+\frac{\Delta}{2})}{\Gamma\left(\Delta - \frac{1+\ann}{2}\right)^2}\,,
\end{equation}  
from which \eqref{eq:Nkgeneric} follows.

The thermal two-point function has an interesting structure as the ratio of functions that are characterized by the dimensional $\Delta$ and the shadow dimension $\tilde{\Delta}$. While this was known for the minimally coupled fields in the BTZ background (see \cite{Jana:2020vyx}), it also appears to hold for the thermal correlators in higher dimensions \cite{Dodelson:2022yvn}.

A similar analysis can be carried out if we quantize the field with Neumann boundary conditions when $\Delta < \frac{1+\ann}{2}$. In this case we fix the conjugate momentum as the source \eqref{eq:sourceJalt}. As noted at the end of \cref{sec:BTZtoy} ingoing boundary-bulk Green's function for this can be obtained from the result for the Dirichlet boundary condition \eqref{eq:Ginbtz} by the replacement $\Delta \leftrightarrow 1+\ann - \Delta$. This implies that  the retarded Green's functions for Dirichlet and Neumann boundary conditions are inverses of each other viz.,\footnote{
   We are normalizing the sources for the Dirichlet and Neumann boundary conditions with a slight difference to enable this simple inversion of the retarded propagator. 
}
\begin{equation}\label{eq:Kinrelns}
\frac{1}{2\left(\Delta - \frac{1+\ann}{2}\right)}\, \Kin{\ann,\Delta}(k)  = \frac{2 \left(\tilde{\Delta} -\frac{1+\ann}{2}\right)}{\Kin{\ann,\tilde{\Delta}}(k) } \,. 
\end{equation} 
%

\subsection{Analytic structure of the retarded correlator and stability}
\label{sec:astab}

We now analyze the analytic structure of the boundary retarded correlator uncovering a surprising feature. There exists a region of the $(\ann,\Delta)$ parameter space in which the BTZ black hole has a linear instability! This is somewhat curious for one would have thought that being a quotient of \AdS{3} the solution is linearly stable towards perturbation by designer scalar probes. The change in the boundary conditions however appears to have a strong effect. A similar phenomenon for minimally coupled scalars with Robin boundary conditions was discovered in \cite{Dappiaggi:2017pbe}.    

\paragraph{Dirichlet quantization:} The retarded two-point functions have a series of simple poles inherited  from the $\Gamma$ functions in $\gfn{\ann}(k,\Delta)$. Owing to the rational form of the expression we find a set of at frequencies $\bwt_{_\text{D}}$ conditioned as follows\footnote{
   We have placed a subscript D to remind us that the field was quantized with Dirichlet boundary conditions.}
\begin{equation}\label{eq:qnormalcond}
\left\{\bwt_{_\text{D}} \,\bigg| \bpt_+ + \frac{\Delta}{2} = - n_+ \;\;\text{or}\;\;  \bpt_- + \frac{\Delta}{2} = - n_- \;\;  \text{and} \;\; \left(\bpt_\pm +\frac{\tilde{\Delta}}{2} \neq - m_\pm\right) \,, \;\; n_\pm, m_\pm \in \mathbb{Z}_{\geq0}\,\right\} .
\end{equation} 
That is, we pick up the only those poles from the Gamma functions appearing in  the numerator of \eqref{eq:Kindef}, which are not simultaneously zeros of the denominator. Recalling that our solution is valid for generic values of $\ann$ and $\Delta$ satisfying  $\Delta-\tilde{\Delta} \neq \mathbb{Z}$, we find a set of discrete poles of the two-point retarded correlator for generic $\bqt$ at   
\begin{equation}\label{eq:Dqnms}
\bwt_{_\text{D}} = \pm\, \sqrt{\bqt^2 - \frac{(\ann-1)^2}{4}} - i\, \left(\Delta + 2 \, n - \frac{\ann-1}{2}\right) \,,\qquad n \in \mathbb{Z}_{\geq 0} \,.
\end{equation} 
Taking $\bwt$ and $\bqt$ to be complex-valued, the above defines the quasinormal spectral curve. At special discrete points on this curve, however, the correlator is analytic. For the present case, this happens at the loci $(\bwt_{_\text{D}}^\ast, \bqt_\ast)$ characterized by $ m,n \in \mathbb{Z}_{\geq 0}$
\begin{equation}\label{eq:DqnmsA}
\begin{split}
\bwt_{_\text{D}}^\ast &= 
    -i \, (1+ n+m) \,, \\ 
\bqt_\ast &= 
   \pm i\, \sqrt{ \ann + (\Delta + n - m)(\ann+1+\Delta+n-m)} \,.
\end{split}
\end{equation} 
We will refer to this set of points on the quasinormal spectral curve as \emph{apparent quasinormal modes}.\footnote{
   In the literature this phenomenon is often referred to as the `pole-skipping' behaviour of the thermal Green's function. We find the terminology confusing when applied to $\Kin{\ann,\Delta}(k)$ since there is no pole to be skipped -- the correlator is analytic at these loci. We will return to this point later in \cref{sec:discuss}. \label{fn:nopolesareskipped}
} 
Note that this phenomenon is already present for thermal correlators of quasi-primaries in a two-dimensional CFT as discussed in \cite{Grozdanov:2019uhi,Blake:2019otz}. Usually, in discussions of quasinormal modes, one takes the momenta to be real, as one is interested in late-time decay of linearized perturbations (which are superpositions of the quasinormal mode functions). The analytic properties of the retarded two-point function, however, are naturally discussed with both $\bwt, \bqt \in \mathbb{C}$.

For our model to be physical, the poles of the retarded Green's functions should lie in the lower half of the complex frequency plane. Since $\Delta > \frac{\ann+1}{2}$ with this choice of boundary condition, stability is guaranteed at high momenta $\bqt$ where the quantity inside the square root is positive. For low momenta, we expand out the answer in powers of $\bqt^2$, obtaining
\begin{equation}\label{eq:Dqnmgap}
\bwt_{_\text{D}} = 
\begin{cases}
&  - i\, \left(\Delta -\ann+1 + 2\, n\right)  -\frac{i\, \bqt^2}{\ann-1} + \order{\bqt^4}\,, \qquad \ann > 1 \,,\\
& - i\, \left(\Delta  + 2\, n\right)  -\frac{i\, \bqt^2}{1-\ann}  + \order{\bqt^4} \,, \qquad  \qquad \qquad\ann < 1 \,.
\end{cases}
\end{equation} 
There are no apparent quasinormal modes for low momenta, as all modes at these discrete loci have $\bwt \sim \order{1}$, i.e., the frequencies are of order the thermal scale.

The poles of the retarded correlator lie on the lower half-plane provided
\begin{itemize}[wide,left=0pt]
\item  $\Delta > \max\{\ann-1, \frac{1+\ann}{2}\}$ with $\ann >1$, or
\item $\Delta \geq \max\{0, \frac{\ann+1}{2}\}$ with $\ann < 1$. 
\end{itemize}
Note, in particular, that the minimally coupled case $\ann =1$ is always stable for scalar primary operators satisfying unitary bound $\Delta >0$. 

Interestingly, for $\ann\in (-\infty,-1)\cup(3,\infty)$ there is a choice of $\Delta$ for which the $n=0$ mode has diffusive dispersion relation. This is the characteristic feature of non-Markovian behaviour. In this situation the Green's function does not have an analytic low energy expansion, owing to the presence of this ungapped mode.

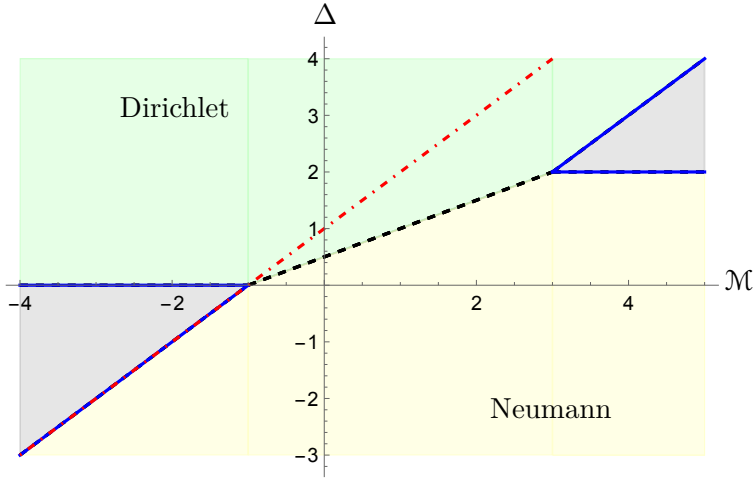
\begin{figure}[th!]
\begin{center}
\begin{tikzpicture}[scale=0.5]
\node (img) {\includegraphics[scale=0.5]{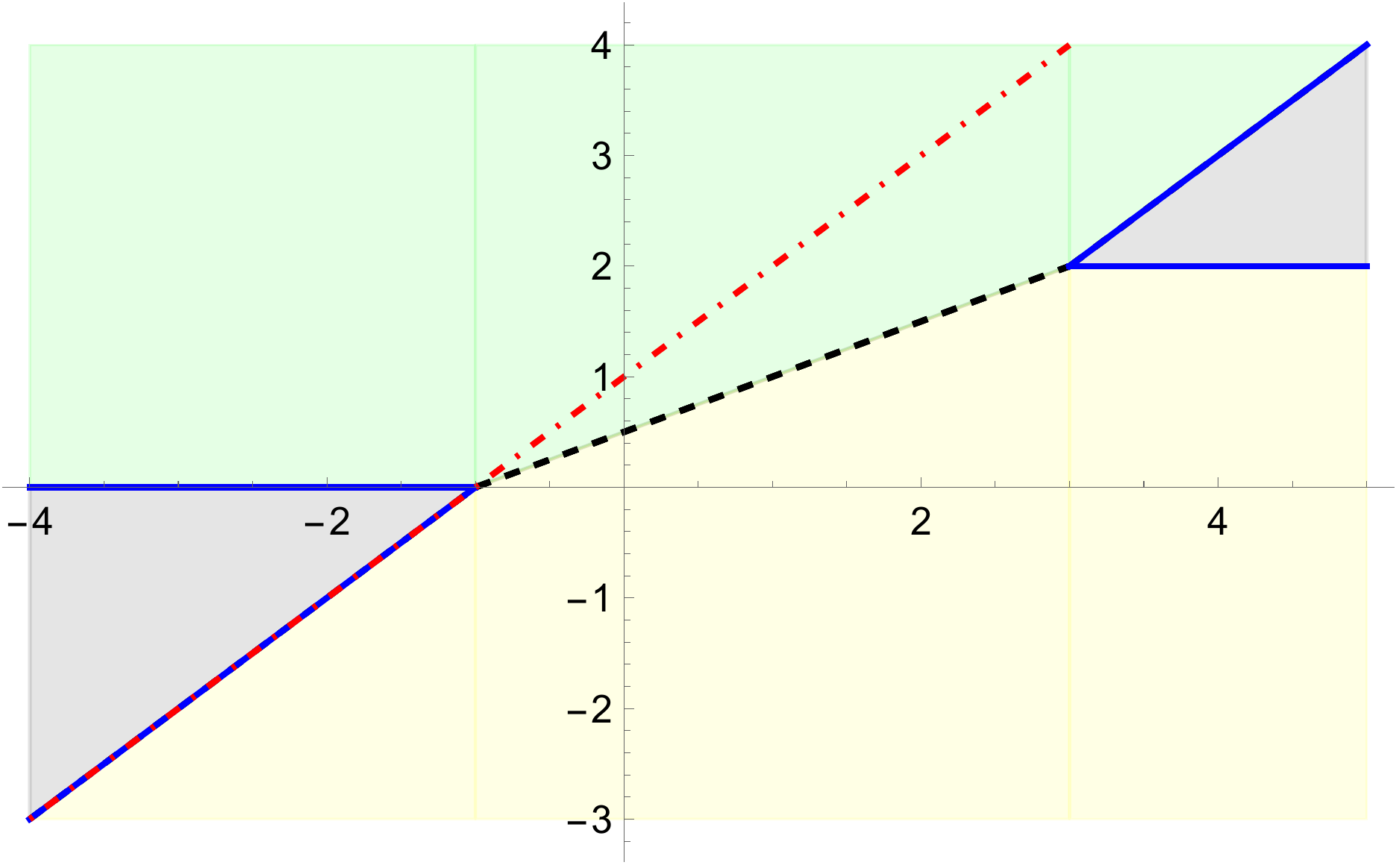}};
\node at (10,-0.6) {$\ann$};
\node at (-1,6.5) {$\Delta$};
\node at (-5,4) {Dirichlet};
\node at (5,-4) {Neumann};
\end{tikzpicture}
\caption{ We illustrate the features of the models we consider in the two-dimensional $(\ann,\Delta)$ parameter space. The Dirichlet and Neumann boundary conditions are separated across the locus $\Delta = \frac{1+\ann}{2}$, indicated by the dashed black line. The gray regions correspond to domains where the lowest quasinormal mode is unstable at low momenta. At its boundary indicated by the blue lines, we encounter the existence of diffusive modes, analogous to the hydrodynamic modes encountered for higher dimensional black holes. We have also indicated the massless case, whence $\Delta=\ann+1$, which we note is always stable in the Dirichlet regime and has hydrodynamic behaviour in the Neumann regime.}
\label{fig:DNDiag}
\end{center}
\end{figure}

It is curious to see the appearance of non-Markovian operators with Dirichlet boundary conditions. All the examples encountered in higher dimensions in \cite{Ghosh:2020lel,He:2022jnc,He:2021jna,He:2022deg} the non-Markovian operators were massless and quantized with Neumann boundary conditions (for computing correlators). However, once we allow for massive fields, we see that the relative independence of the conformal dimension $\Delta$ and the dialtonic index $\ann$ allows for a more intricate interplay. We illustrate the general features of the boundary conditions, stability and presence of gapless modes in the $(\ann,\Delta)$ parameter space in \cref{fig:DNDiag}.

Since we have an exact expression  we can continue to work with the correlators even in the presence of an ungapped mode. However, should we be interested in having  low energy effective description, then this would not suffice. In that case, as advocated in \cite{Ghosh:2020lel}, rather than deriving the generating functional of correlation functions parameterized by the sources $\{J_\skR, J_\skL\}$, one computes a Legendre transformed object, the Wilsonian influence functional, parameterized by the operator expectation values 
$\{\snMar_\skR, \snMar_\skL \}$.  

For the operator $\mathcal{O}$, whose dual field $\sen{}$ satisfies Dirichlet boundary conditions,  the field values $\snMar$ are given in terms of the conjugate momentum
\begin{equation}\label{eq:nMopD}
\snMar = \expval{\mathcal{O}} =  - \frac{1}{\Delta - 1 - \ann}\,  \lim_{r\to \infty} 
  \, \left(\frac{1}{r^{\Delta}}\, \pi + \text{counterterms}\right).
\end{equation} 
The Wilsonian influence functional can therefore be computed by quantizing the field with Neumann boundary conditions instead. All the Legendre transform does is invert the Green's function, so the Wilsonian effective action for non-Markovian fields reads
\begin{equation}\label{eq:DnMwif}
S_{\ann,\Delta}^{(2)} [\snMar_a,\snMar_d]
= -\frac{1}{2} \, \int_k \, \snMar_d(-k)\, 
   \frac{1}{\Kin{\ann,\Delta}(k)}\,  
   \left[\snMar_a(k) + \left(\nB(\omega)+\frac{1}{2}\right)\, \snMar_d(k) \right] ,
\end{equation} 

We will for the most part focus on computing the generating function of correlators in what follows. However, once we have the result, we will explain the salient features which we expect for the Wilsonian influence functionals. The analytic control of our model makes it quite straightforward to translate between the two pictures.\footnote{
   In a certain sense this is analogous to the observation that non-Markovian data for massless fields with index $\ann <-1$ can be obtained from that for Markovian fields with index $\ann >1$ by analytically continuing the index to negative values \cite{Ghosh:2020lel,He:2021jna}.
}

\paragraph{Neumann quantization:} The above discussion can be generalized immediately to the case where we quantize the fields with Neumann boundary conditions. The quadratic generating function of correlators can be immediately written down

\begin{equation}\label{eq:S2adN}
S_{\ann,\Delta}^{(2)} [\widehat{J}_a,\widehat{J}_d]
= -\frac{1}{2} \, \int_k \, \widehat{J}_d(-\bwt,-\bqt)\, \Kin{\ann,\tilde{\Delta}}(\bwt,\bqt)\,  
   \left[\widehat{J}_a(\bwt,\bqt) + \left(\nB+\frac{1}{2}\right)\, \widehat{J}_d(\bwt,\bqt) \right] .
\end{equation} 
While the result appears to be similar to the Wilsonian influence functional for operators quantized with Dirichlet boundary condition owing to \eqref{eq:Kinrelns}, we emphasize that the two expressions \eqref{eq:DnMwif} and \eqref{eq:S2adN} are qualitatively different.  The quasinormal poles for the retarded Green's function (subscript N) are correspondingly located for generic momenta at 
\begin{equation}\label{eq:Nqnm}
\bwt_{_\text{N}} = \pm\, \sqrt{\bqt^2 - \frac{(\ann-1)^2}{4}} - i\, \left(-\Delta + 2 \, n + \frac{\ann+3}{2}\right) \,, \qquad n \in 
   \mathbb{Z}_{\geq 0} \,.
\end{equation} 
The apparent quasinormal modes are present at the discrete points $(\bwt_{_\text{N}}^\ast, \bqt_\ast)$ 
\begin{equation}\label{eq:DqnmsA}
\begin{split}
\bwt_{_\text{N}}^\ast &= 
    -i \, (1+ n+m) \,, \\ 
\bqt_\ast &= 
   \pm i\, \sqrt{ \ann + (\tilde{\Delta} + n - m)(\ann+1+\tilde{\Delta}+n-m)} \,,
\end{split}
\end{equation} 
with $ m,n \in \mathbb{Z}_{\geq 0}$.

Since $\Delta < \frac{\ann+1}{2}$ for operators quantized with Neumann boundary condition, stability is guaranteed for high momenta $\bqt > \abs{\frac{\ann-1}{2}}$. For small momenta, on the other hand, we have 
\begin{equation}\label{eq:Dqnmgap}
\bwt_{_\text{N}} = 
\begin{cases}
&  - i\, \left(2-\Delta+ 2\, n\right)  -\frac{i\, \bqt^2}{\ann-1} + \order{\bqt^4}\,, \qquad\qquad  \ann > 1\,, \\
& - i\, \left(\ann+1-\Delta  + 2\, n\right)  -\frac{i\, \bqt^2}{1-\ann}  + \order{\bqt^4} \,,  \qquad\ann < 1 \,.
\end{cases}
\end{equation} 
Therefore, the model is stable to linear perturbations only when 
\begin{itemize}[wide,left=0pt]
\item  $\Delta < \min\{2,\frac{\ann+1}{2}\}$ with $\ann >1$, or
\item $\Delta < \min\{\ann+1, \frac{\ann+1}{2}\}$ with $\ann < 1$. 
\end{itemize}
The stability domain can pictorially read-off from \cref{fig:DNDiag}.

Once again it is possible to choose $\Delta \in (-\infty,-1) \cup (3,\infty)$ such that the $n=0$ quasinormal mode is long-lived with diffusive dispersion.  In particular, massless fields with $\ann <-1$ always have a non-Markovian mode. This is exactly the class of designer scalars that have been encountered in the higher dimensional black hole context. While the index in those cases is integral, we will refrain from making that choice to respect the genericity condition $\Delta - \tilde{\Delta} \notin \mathbb{Z}$.\footnote{
   Some particular cases violating the genericity condition have been analyzed in the literature, eg., \cite{Son:2002sd} studied the behaviour of primary correlators in two-dimensional CFTs with integral dimensions, and more recently \cite{Grozdanov:2019uhi} examined $U(1)$ current correlators in a three-dimensional model with translational symmetry breaking described at the end of \cref{sec:setup}. In these cases, the boundary Green's function is not of the rational form, but is given in terms of digamma functions.}

In the presence of a gapless quasinormal mode, we can construct the Wilsonian influence functional by Legendre transforming the generating function. For fields with $\Delta < \frac{1+\ann}{2}$, which quantized with Neumann boundary conditions, this amounts to instead quantizing with renormalized Dirichlet boundary conditions. For the Gaussian effective action the effect is to invert the kernel in \eqref{eq:S2adN}. This has been extensively discussed in the higher dimensional examples, where the only analytic expressions available are in a low energy gradient expansion.

\subsection{Non-Gaussian influence functionals}
\label{sec:nongauss}

We have thus far computed the ingredients that enter the open effective action at the quadratic order. Our next step is to consider the cubic interaction term in \eqref{eq:Sdesign} and compute the non-Gaussian corrections. Note that we have posited that the bulk cubic vertex is itself modulated by a dilaton $\chi_\lambda$. For simplicity, we will take this to be a power law as well, letting 
the cubic vertex be
\begin{equation}\label{eq:QubicP}
S_{(3)}[\sen{1},\sen{2},\sen{3}] = - \lambda\,  \oint d^2x\,d\ctor\, \sqrt{-g} 
   \left(\frac{r}{r_+}\right)^{\ann_\lambda} \, \sen{1}\,\sen{2}\,\sen{3} \,.
\end{equation}

The cubic vertex leads at leading order to two types of contributions to the effective action: cubic terms which arise from a bulk contact diagrams, and quartic contributions, arising from a bulk exchange diagram. We are going to assume for simplicity that the three fields interacting at the vertex have been quantized with Dirichlet boundary conditions. As noted above the switch to Neumann boundary conditions can be easily achieved by replacing the operator dimension by the shadow dimension. 

\subsubsection{Cubic influence functional}
\label{sec:cubic}

The contact diagram for a bulk cubic interaction has already been evaluated in \cite{Jana:2020vyx} for the case of a minimally coupled scalar field in the BTZ geometry (whence $\ann_{1,2,3}=1$ and $\ann_\lambda =0$). Allowing for non-trivial dilatons does not change the qualitative nature of the bulk integrals to evaluate, so we will be brief, mainly quoting the result in what follows. 

By the Schwinger-Keldysh and KMS conditions the FFF and PPP correlators vanish in the retarded-advanced basis. Furthermore, FFP and PPF are complex conjugates of each other. Thus, we only have one diagram to compute. Using \eqref{eqn:Icontact} we find that we need to the following single-sheet integral 
\small
\begin{equation}\label{eq:FFPS}
\begin{split}
\mathcal{I}^{123}_{_\text{FFP}}(k_1,k_2,k_3) 
&= 
   -\lambda\, \left(1-e^{2\pi\,\bwt_3}\right)
      \int_0^{1}\, \frac{dz}{z^{\ann_\lambda+3}}
         \left(\frac{1-z}{1+z}\right)^{i\,\bwt_3} \\
&\qquad \qquad\qquad \qquad  \times
         \Gin{\ann_1,\Delta_1}(z,k_1)\, \Gin{\ann_2,\Delta_2}(z,k_2) \, \Gin{\ann_3,\Delta_3}(z,\wk_3)
\,.
\end{split}
\end{equation}
\normalsize

The ingoing boundary-bulk Green's function in \eqref{eq:Ginbtz} can be conveniently expressed as an integral using the Barnes' representation of the hypergeometric function as 
\small
\begin{equation}\label{eq:BtBBarnes}
\Gin{}(z,k) = 
   \frac{z^{\Delta}(1+z)^{-i\,\bwt}}{\gfn{\ann}(k,\tilde{\Delta})}
   \, 
      \int_{\mathcal{C}}\, \frac{ds}{2\pi i}\,
         \Gamma(s)\, \Gamma\left(\tilde{\Delta}-\frac{\ann+1}{2}+s\right) \, 
         \Gamma\left(\bpt_+ +\frac{\Delta}{2}-s\right)\,
         \Gamma\left(\bpt_-+\frac{\Delta}{2}-s\right) \, z^{-2s}\,,
\end{equation}
\normalsize
The contour $\mathcal{C}$ runs parallel to the imaginary axis and is chosen to as to separate the poles at $s=-n$ and $s=\frac{\ann+1}{2}-\tilde{\Delta}-n$ from those at $s=\bpt_{\pm}+n$. 

We write each of the three boundary-bulk propagators by this integral representation. The radial $z$ integral turns out to converge provided 
\begin{equation}\label{eq:ConvC}
\Re{\sum_{j=1}^3\left(2s_i-\Delta_{a_i}\right) } +\ann_\lambda<-2\,.
\end{equation}
We assume this to hold and complete the radial integral, which gives us a ratio of Gamma functions. The last step is to carry out the three contour integrals from the Barnes' representation. This can be done by closing the contours to the left picking up the poles at $s_i = -n_i$ and $s_i = \Delta_{a_i} -\frac{1+\ann_i}{2} -n_i $, respectively. Most of the sums can be carried out and the final answer as a single sum over a product of generalized hypergeometric functions as in \cite{Jana:2020vyx}. Factoring out the boundary Green's functions, we can express the result as follows
\begin{equation}\label{eq:3contact}
\begin{split}
\mathcal{I}^{123}_{_\text{FFP}}(k_1,k_2,k_3) 
&=
   -\lambda\, \left(1-e^{2\pi\,\bwt_3}\right) \, \Gamma\left(1+i\bwt_3\right) \, \Kin{\ann_1,\Delta_1}(k_1)\, \Kin{\ann_2,\Delta_2}(k_2)\, 
   \Kin{\ann_3,\Delta_3}(\wk_3) \; \\
&\qquad 
   \times 
    \sum_{\delta_i \in \{\Delta_{a_i},\tilde{\Delta}_i\}} \,
   \sum_{n=0}^\infty \; \mathfrak{J}^\delta_{3c}(n) \,, 
\end{split}
\end{equation}
We define here a function $\mathfrak{J}^\delta$, which controls the residues at the poles. The notation is as follows: the parameter $\delta$ can either be the dimension or the shadow dimension for each of the three external operators. The result is a sum over eight choices indicated by the summation in the second line. The function $\mathfrak{J}^\delta$ is itself given as 

\begin{equation}\label{eq:JFFP}
\begin{split}
\mathfrak{J}^\delta_{3c}(n)
=
   \left(\prod_{i=1}^3\, \frac{1}{2\left(\Delta_i - \frac{\ann_i+1}{2}\right)}\right)
    \, \frac{\gfn{\ann_1}(k_1,\delta_1)}{\gfn{\ann_1}(k_1, \Delta_1)}\,\frac{\, \gfn{\ann_2}(k_2,\delta_2)}{\gfn{\ann_2}(k_2,\Delta_2)} \,   
   \frac{\gfn{\ann_3}(\wk_3, \delta_3+2n)}{\gfn{\ann_3}(\wk_3,\Delta_3)} \;
      \mathfrak{H}_n \, \mathfrak{K}  \,,
\end{split}
\end{equation}
where we introduced 
\small
\begin{equation}
\begin{split}   
\mathfrak{H}_n
&=
   \frac{(-1)^n \, \Gamma\left(\frac{\ann_3+1}{2}-n-\delta_3\right)}{\Gamma(1+n) \, \Gamma\left(\frac{\ann_3+1}{2}-2\,n-\delta_3\right)}\, 
   \Gamma\left(\frac{\delta_1+\delta_2+\delta_3}{2} +n-\frac{\ann_\lambda+2}{2}\right)
    \Gamma\left(\delta_1-\frac{\ann_1-1}{2}\right) \,,\\
\mathfrak{K}
&=   
   \pFq{4}{3}{\bpt_{+2}+\frac{\delta_2}{2},\bpt_{-2}+\frac{\delta_2}{2},\frac{1+\ann_3}{2}-n-\delta_3,-n}{1-\bpts_{+3}-n,1-\bpts_{-3}-n,\delta_2-\frac{\ann_2-1}{2}}{1}   \\
&\qquad \qquad    
    \times \;
    \pFqReg{3}{2}{\bpt_{+1}+\frac{\delta_1}{2},\bpt_{-1}+\frac{\delta_1}{2},\frac{\delta_1+\delta_2+\delta_3}{2}+n-\frac{\ann_\lambda+2}{2}}{i\,\bwt_3+\frac{\delta_1+\delta_2+\delta_3}{2} -\frac{\ann_\lambda}{2}+n, \delta_1-\frac{\ann_1-1}{2}}{1} \,.
\end{split}
\end{equation}

When any of the $\delta_i = \Delta_i$, this $\mathfrak{J}^\delta_{_\text{FFP}}$ function is manifestly regular in the corresponding 3-momentum variable. For $\delta_i = \tilde{\Delta}_i$ we do find poles, but these are spurious, as they cancel against the factors of $\Kin{\ann_i,\Delta_i}$ that we have factored out. As promised, the singularities in the frequency domain, can be read off from the boundary retarded Green's function alone. We find that the $\mathcal{I}_{_\text{FFP}}(k_1,k_2,k_3)$ has poles in the lower-half complex $\omega_1$ and $\omega_2 $ planes corresponding to the quasinormal modes, and in the upper half  $\omega_3$ plane corresponding to the anti-quasinormal modes.  This verifies in a special case, the general observations made  in \cref{sec:analcorr}. We will return to extracting the physical content of this expression after recording the four-point function.

\subsubsection{Quartic influence functional}
\label{sec:quartic}

Let us now turn to the computation of four-point function involving one bulk-bulk propagator. To keep the discussion general, we will imagine all the four-external operators are distinct with parameters $(\ann_{a_i},\Delta_{a_i})$, for $i=1,\ldots,4$. We can subsequently simplify these expressions to the case where some operators are equal. 

Accounting for the Schwinger-Keldysh and KMS conditions, which imply the vanishing of FFFF and PPPP correlators, we are left with computing the FFPP, FPFP, and FPPP correlators. We are distinguishing the FFPP and FPFP correlators for the present since we are assuming that the operators are distinct. From the discussion in  \cref{sec:Wittendiag} it suffices to quote the master integral \cref{eq:4ptmaster} for our model, as all 4-point orderings can be recovered from it. For completeness, let us record one of the integrands that enters 
the double bulk-integral \eqref{eq:4pt1exfull}, say for the FFPP ordering
\small
\begin{equation}\label{eq:FFPPintegral}
\begin{split}
\mathcal{I}_{_\text{FFPP}}^{a_1 a_2 a_3 a_4}(k_1,k_2,k_3,k_4) 
&=  
   \mathscr{N}(k)\, \int_{\ctor_c}^{\ctor_+}\, d\ctor\, \int_{\ctor_c}^{\ctor_+}\, d\ctor' \; 
   \Gin{a_1}(k_1,\ctor)\, \Gin{a_2}(k_2,\ctor) \, \Grev{a_3}(k_3,\ctor')\, \Grev{a_4}(k_4,\ctor') \\
&\qquad
   \times \bigg[
         e^{2\pi\,\bwt (1-\ctor)} \, \Grev{e}(\ctor,k) \left[
         \Gin{e}(\ctor',k) - e^{-2\pi\,\bwt} \, \Grev{e}(\ctor',k)\right] \theta(\ctor-\ctor') \\
&\qquad\quad 
    +  e^{2\pi\,\bwt (1-\ctor')} \, \Grev{e}(\ctor',k) \left[
         \Gin{e}(\ctor,k) - e^{-2\pi\,\bwt} \, \Grev{e}(\ctor,k)\right] \theta(\ctor'-\ctor) \bigg] \,.
\end{split}
\end{equation} 
\normalsize
In writing this expression we have introduced the $2$-momentum $k = k_3 + k_4$ (and hence $\bwt = \bwt_3+\bwt_4$, and $\bqt = \bqt_3+\bqt_4$, respectively). We can still use momentum conservation to eliminate $\bwt$ (since $k_1+k_2+k_3+k_4 = 0$) and write the expression as a function of $k_1,k_2, k_3$ alone, but will refrain from doing so. The reader can directly verify that one reproduces the   result quoted in  \eqref{eq:FFPP}.

Thus, we are left with evaluating the master integral \eqref{eq:4ptmaster}. We use again the  Barnes' representation ingoing boundary-bulk propagator \eqref{eq:BtBBarnes}, and write the master integral as the following nested single-sheet integral:
\small
\begin{equation}\label{eq:IMgenbtz}
\begin{split}
\IM{a_1 a_2 a_3 a_4 e}{i \, j}{k_1,k_2,k_3,k_4}{k_5,k_6,k_7,k_8} 
&=
   \left[\prod_{i=1}^{6}\int_{\mathcal{C}_i}\frac{ds_i}{2\pi i}
   \frac{\Gamma(s_i)\,\Gamma\left(s_i + \frac{1+\ann_{a_i}}{2}-\Delta_{a_i}\right)}{\Gamma\left(\frac{1+\ann_{a_i}}{2}-\Delta_{a_i}+2\, s_i\right)}
   \frac{\gfn{\ann_{a_i}}(k_i,\Delta_{a_i}-2\,s_i)}{\gfn{\ann_{a_i}}(k_i,\tilde{\Delta}_{a_i})}\right]\\ 
&\quad 
   \times 
   \int_{0}^{1}\, dz
      \left(\frac{1+z}{1-z}\right)^{i\bwt_7}
      (1+z)^{-i (\bwt_1+\bwt_2+\bwt_5)}\, 
      z^{\alpha_1}\\ 
&\quad 
   \times
   \int_{0}^{z}\, dz'
   \left(\frac{1+z'}{1-z'}\right)^{i\bwt_8}
      (1+z')^{-i (\bwt_3+\bwt_4+\bwt_6)}\,
         z'^{\alpha_2} \,,
\end{split}
\end{equation}
\normalsize 
where we defined the exponents\footnote{
   We have introduced $\ann_{a_5} = \ann_{a_6} = \ann_e$ and $\Delta_{a_5} = \Delta_{a_6}= \Delta_6$ to write the expressions compactly. 
}
\begin{equation}\label{eq:alphasdef}
\alpha_1 = - \ann_{\lambda_i}-3 + \sum_{m\in\{1,2,5\}} \, (\Delta_{a_m}-2\, s_m)
\,, \qquad 
\alpha_2 = - \ann_{\lambda_j}-3 + \sum_{m\in\{3,4,6\}} \, (\Delta_{a_m}-2\, s_m)
\,. 
\end{equation} 

This general expression turns out to be hard to evaluate. The inner integral can be completed in terms of Appell functions, but that leaves us with a complicated outer integral. However, for the computation of the influence functionals we do not need the expressions in full generality. As noted in \cref{sec:Wittendiag} it suffices to analyze the integral subject to the constraints \eqref{eq:om78constrained}. With this assumption, \eqref{eq:IMgenbtz} simplifies considerably. Focusing on the radial integrals we need to evaluate 
\begin{equation}\label{eq:IMbtzradial}
\int_{0}^{1}\, dz 
   \left(1-z^2\right)^{-i\bwt_7}\,z^{\alpha_1} 
   \int_{0}^{z}\, dz' 
   \left(1-z'^2\right)^{-i\bwt_8}\,z'^{\alpha_2}\,.
\end{equation} 
These nested integrals can be done in terms of generalized hypergeometric functions. Note that the convergence of the integral demands a constraint on $\alpha_{1,2}$
\begin{equation}\label{eq:4ptconverge}
\Re(\alpha_1+\alpha_2) > -2 \;\;\longrightarrow \;\; 
\Re{\sum_{i=1}^{6}\left(2\,s_i-\Delta_{a_i}\right)}+ \ann_{\lambda_i}+ \ann_{\lambda_j}<-4\,,
\end{equation}
as well as one on the frequencies:\footnote{
   These are the conditions for the hypergeometric sum to converge at the branch point $z=1$, \cite[16.2.2]{NIST:DLMF}.
}
\begin{equation}\label{eq:4ptw78}
\Im(\bwt_7) >-1 \,, \qquad \Im(\bwt_7 + \bwt_8) > -2 \,.
\end{equation} 
We will assume this to hold for the present. With its aid, we can show that the master integral reduces to the following set of contour integrals
\small
\begin{equation}\label{eq:IMbtzcontour}
\begin{split}
&\IM{a_1 a_2 a_3 a_4 e}{i \, j}{k_1,k_2,k_3,k_4}{k_5,k_6,k_7,k_8} \\
&=
   \frac{1}{4}\Gamma\left(1-i\bwt_7\right) 
   \left[\prod_{i=1}^{6}\int_{\mathcal{C}_i}\frac{ds_i}{2\pi i}
   \frac{\Gamma(s_i)\,\Gamma\left(s_i + \frac{1+\ann_i}{2}-\Delta_{a_i}\right)}{\Gamma\left(\frac{1+\ann_i}{2}-\Delta_{a_i}+2\,s_i\right)} \, 
   \frac{\gfn{\ann_i}(k_i,\Delta_{a_i}-2\,s_i)}{\gfn{\ann_i}(k_i,\Delta_{a_i})}\right]
      \\
&\qquad 
\times 
    \Gamma\left(\frac{1+\alpha_2}{2} \right)\, \Gamma\left(\frac{\alpha_1+\alpha_2}{2} +1\right)\,
   \pFqReg{3}{2}{i\bwt_8, \frac{\alpha_2+1}{2}  ,\frac{\alpha_1+\alpha_2+2}{2}}{\frac{\alpha_2+3}{2},-i\bwt_7+\frac{\alpha_1+\alpha_2}{2}+2}{1} \,.
\end{split}
\end{equation}
\normalsize 
Note that the parameters $\alpha_1$ and $\alpha_2$ depend on the $s_i$ which are integrated over. The convergence condition \eqref{eq:4ptconverge} is analogous to the one for contact diagrams \eqref{eq:ConvC} and constrains the external operator dimensions. The constraints on the frequencies \eqref{eq:4ptw78} simply is a statement of causality; one can check that it requires the correlator to follow the analyticity properties dictated by the F and P labels. 

The contour integrals can be done as before using residue calculus, and the expression written as a sixfold sum. We are able to complete two of the sums in terms of hypergeometric functions and record here the final answer as a fourfold sum over $n_i$ with $i=1,\cdots 4$. 
\small
\begin{equation}\label{eq:4ptmasterF}
\begin{split}
\IM{a_1 a_2 a_3 a_4 e}{i \, j}{k_1,k_2,k_3,k_4}{k_5,k_6,k_7,k_8} 
&=
   \frac{1}{4}\,\Gamma\left(1-i\bwt_7\right)\, \left(\prod_{i=1}^6\,
         \Kin{\ann_{a_i}, \Delta_{a_i}}(k_i)\right)\, \\
& \qquad \times 
   \sum_{\delta_i=\{\Delta_{a_i},\tilde{\Delta}_{a_i}\}}
   \sum_{n_1,n_2,n_3,n_4=0}^\infty\, \mathfrak{J}^\delta_{4ex}(n_1,n_2,n_3,n_4) \,.\end{split}
\end{equation}
\normalsize
We have once again factored out the essential pieces involving the boundary retarded Green's functions. Now we not only have the factors corresponding to the external operators, but also additional contributions which depends on the internal bulk exchange, the $k_5$ and $k_6$ terms. We recall that these momenta take values $k$ or $\wk$.  In the actual influence functionals there is an additional factor of $\mathscr{N}_{\ann_e,\Delta_e}(k)$, the normalization factor of the bulk-bulk propagator \eqref{eq:Nfix}, which cancels some poles from these additional factors of $\Kin{}$ (ensuring that there are no double poles). There are additionally Matsubara poles from the factor $\Gamma(1-i\bwt_7)$, which we can check always corresponds to the exchanged frequency when it remains uncanceled by the statistical factors.

The function that captures the residue information $\mathfrak{J}^\delta_{4ex}$ is in its turn given by
\scriptsize
\begin{equation}
\begin{split}
&\mathfrak{J}^\delta_{4ex}(n_1,n_2,n_3,n_4)\\
&= 
   \frac{\gfn{\ann_5}(k_5,\delta_5)}{\gfn{\ann_5}(k_5,\Delta_5)}
   \frac{\gfn{\ann_6}(k_6,\delta_6)}{\gfn{\ann_6}(k_6,\Delta_6)}    
   \prod_{i=1}^{4}\frac{(-1)^{n_i}}{\Gamma(1+n_i)}
   \frac{\Gamma\left(\frac{1+\ann_i}{2}-\delta_i-n_i\right)}{\Gamma\left(\frac{1+\ann_i}{2}-\delta_i-2n_i\right)}\, 
    \frac{\gfn{\ann_i}(k_i,\delta_i+2n_i)}{2\,\left(\Delta_i-\frac{1+\ann_i}{2}\right) \, \gfn{\ann_i}(k_i,\Delta_{a_i})} \\ 
&\quad 
   \times 
   \Gamma\left(\gamma_1^\delta-1+n_1+n_2\right)
   \Gamma\left(\gamma_2^\delta-2+\sum_{i=1}^{4}n_i\right)
   \pFqReg{3}{2}{\gamma_1^\delta-1+n_1+n_2,\gamma_2^\delta-2+\sum_{i=1}^{4}n_i,i\bwt_8}{\gamma_1^\delta+n_1+n_2,-i\bwt_7+\gamma_2^\delta-1+\sum_{i=1}^{4}n_i}{1}\\ 
&\qquad 
   \times 
   \pFq{4}{3}{\bpt_{+5}+\frac{\delta_5}{2},\bpt_{-5}+\frac{\delta_5}{2},\frac{1+\ann_2}{2}-\delta_2-n_2,-n_2}{1-\bpt_{+2}-\frac{\delta_2}{2}-n_2,1-\bpt_{-2}-\frac{\delta_2}{2}-n_2,\frac{1-\ann_5}{2}+\delta_5}{1}\\ 
&\quad \qquad
\times 
\pFq{4}{3}{\bpt_{+6}+\frac{\delta_6}{2},\bpt_{-6}+\frac{\delta_6}{2},\frac{1+\ann_4}{2}-\delta_4-n_4,-n_4}{1-\bpt_{+4}-\frac{\delta_4}{2}-n_4,1-\bpt_{-4}-\frac{\delta_4}{2}-n_4,\frac{1-\ann_6}{2}+\delta_6}{1}
\,.
\end{split}
\end{equation}
\normalsize 
We have defined
\begin{equation}\label{eq:}
\gamma_1^\delta = \frac{\delta_1+\delta_2+\delta_5- \ann_{\lambda_i}}{2} \,,
\qquad 
\gamma_2^\delta = \frac{1}{2}\sum_{i=1}^{6}\delta_i-\frac{ \ann_{\lambda_i}+ \ann_{\lambda_j}}{2}
\end{equation} 
In carrying out the sums over $n_5$ and $n_6$ (the poles of $s_5$ and $s_6$ contour integrals) we have made some choices for pairing them with the remaining summation variables. 

This completes the derivation of the master integral in terms of which the various 4-point functions are given in \eqref{eq:FFFP}-\eqref{eq:FPFP}. The reader  can check from these expressions that the analytic properties of the correlators delineated at the end of \cref{sec:analcorr} are confirmed by these expressions. 

\section{Physical lessons for open quantum systems}
\label{sec:phylessons}

We have all the necessary ingredients to extract some general lessons for open quantum systems, thanks to the general arguments in \cref{sec:analcorr} and the explicit results in our two-dimensional toy model  \cref{sec:BTZtoy}. We now examine some specific features for both Markovian and non-Markovian modes. To keep the discussion organized, we first explain features when only one of these types of fields is present, and then turn to the case where they interact. 

\subsection{Markovian self-interactions}
\label{sec:Mself}

The simplest case in question is the self-interaction of a set of fields, all of whose modes are short-lived. This is the case for minimally coupled massive scalars. This was already explored in \cite{Jana:2020vyx} correlators computed using contact diagrams. The exchange diagrams do not substantially alter the picture.  

Consider for the sake of simplicity, a single field $\phi$, whose dual operator $\mathcal{O}$ has no long-lived modes. We will also consider only the scalar correlators, and take the field to have a cubic vertex, with coupling $\lambda$ and no vertex function $\chi_\lambda(r)=0$. The bulk Lagrangian is then
\begin{equation}\label{eq:Markov3dmodel}
S[\phi] = 
   -\oint d^2x\,d\ctor\, \sqrt{-g}  
    \left[ 
   \frac{1}{2} \,\left(\frac{r_+}{r}\right)^{\ann-1}\left(\nabla_A \phi \nabla^A\phi + m^2 \, \phi^2\right) +\lambda\, \phi^3\right] .
\end{equation} 

To write the generating functions in a compact form, we introduce the following notation:
\begin{equation}\label{eq:convolprod}
\mathfrak{F}\cdot (X_1, X_2, \cdots , X_n) =
 \left(\prod_{i=1}^n \int_{k_i}\right) \mathfrak{F}(k_1,\cdots, k_n) \, X_1(k_1)\,X_2(k_2)\cdots X_n(k_n) \,\delta(\sum_{j=1}^n\, k_j) \,.
\end{equation} 
With the aid of \eqref{eq:convolprod} the generating function for the correlators takes the following form:
\begin{equation}\label{eq:MarkovS234}
\begin{split}
S_\text{gen}[\JF,\JP] 
&= 
  -\left( S^{(2)}  + S^{(3)} + S^{(4)} \right) \,,\\
S^{(2)}
&=    
   \left(\frac{\JP}{\nB} \right) \cdot (\Kin{\mathcal{O}} \, \JF) 
= \int_{k_1,k_2} \, \frac{\JP(k_1)}{\nB(\omega_1)} \Kin{\mathcal{O}}(k_2)\, \JF(k_2)\, \delta(k_1+k_2) \,,   
\\
S^{(3)}
&= 
   \mathcal{I}_{_\text{FFP}}\cdot(\JF, \JF, \JP) +  \text{F} \leftrightarrow \text{P} \,, \\
S^{(4)}
&=    
   \mathcal{I}_{_\text{FFFP}}\cdot (\JF,\JF, \JF, \JP) + \frac{1}{2}\,\mathcal{I}_{_\text{FFPP}}\cdot (\JF,\JF, \JP, \JP) +  \text{F} \leftrightarrow \text{P} \,.
\end{split}
\end{equation} 
We have singled out the quadratic part of the generating function to make the dependence on the retarded Green's function of the operator $\mathcal{O}$ manifest. In the process, we defined 
\begin{equation}\label{eq:KOdef}
\Kin{\mathcal{O}}(k) \equiv \Kin{\ann,\Delta}(k) \,.
\end{equation} 
The cubic and quartic influence functionals for our model can be read-off from the previous section. For the quartic case, the sum over channels for the bulk exchange has been performed in writing the above, so the functions $\mathcal{I}_{_\text{FFFP}}$ and $\mathcal{I}_{_\text{FFPP}}$ are suitable combinations of the master integrals \eqref{eq:4ptmasterF}. The exchange of F and P labels, can be achieved by frequency reversal on the corresponding $2$-momentum, viz., $k \to \wk$ (for parity even systems).

The reader can confirm the analytic structure of the correlators are as predicted in \cref{sec:analcorr}. In particular, (nb: $k= k_3+k_4$)
\begin{equation}\label{eq:Ipoles}
\begin{split}
\mathcal{I}_{_\text{FFP}}(k_1,k_2,k_3) 
&\propto
    \Kin{\mathcal{O}}(k_1) \, \Kin{\mathcal{O}}(k_2)\,\Kin{\mathcal{O}}(\wk_3) \\
\mathcal{I}_{_\text{FFFP}}(k_1,k_2,k_3,k_4) 
&\propto 
    \Kin{\mathcal{O}}(k_1) \, \Kin{\mathcal{O}}(k_2)\,\Kin{\mathcal{O}}(k_3)\, \Kin{\mathcal{O}}(\wk_4) \left[\mathfrak{R}_0 + \mathfrak{R}_1 \,\Kin{\mathcal{O}}(\wk)  \right]\\
\mathcal{I}_{_\text{FFPP}}(k_1,k_2,k_3,k_4) 
&\propto  
   \Kin{\mathcal{O}}(k_1) \, \Kin{\mathcal{O}}(k_2)\,\Kin{\mathcal{O}}(\wk_3)\, \Kin{\mathcal{O}}(\wk_4) \,\left[\mathfrak{R}_2 + \mathfrak{R}_3 \,\Kin{\mathcal{O}}(k) + \mathfrak{R}_4\, 
    \Kin{\mathcal{O}}(\wk) \right] .
\end{split}
\end{equation} 
The functions $\mathfrak{R}_i$ control the residues -- they have dependence on the momentum labels, which are not indicating. 
What is now manifest is that this is the structure expected from the field theoretic Schwinger-Keldysh and KMS conditions. Real-time diagrammatics, involves only a FP 2-point function and a FFP and PPF 3-point functions. Therefore, the singularities of the 4-point function in the composite momentum $k$ signal the particular channel in which we are decomposing the correlator.

A scalar primary operator of a two-dimensional CFT is a particular example of the type of operator we are considering. In terms of the parameters of the model \eqref{eq:Markov3dmodel}, we set $\ann =1$; in this case $\Delta$ is indeed the conformal dimension. The two-point function \eqref{eq:Kindef} is a well-known result dating back to \cite{Gubser:1997cm} and was derived holographically in \cite{Son:2002sd}. The three-point function was first analyzed in \cite{Becker:2014jla}; they computed the Fourier transform of the Euclidean (cylinder) correlator and expressed it in terms of Mathieu functions. In \cite{Jana:2020vyx} it was computed holographically and expressed in the form quoted in  \eqref{eq:3contact}. The four-point function, as far as we are aware, is new and has not been obtained in the literature before. We describe some further applications of our analysis in this particular context in \cref{sec:discuss}.

This data can now be used to construct the effective action of the open quantum field theory. Consider for simplicity, the quantum system, which we use to probe the holographic environment, to be a free scalar field $\Psi$. We start with the system-environment action (in 2d Minkowski spacetime)
\begin{equation}\label{eq:PsiOact}
S_\text{SE} = - \int d^2x\, \partial_\mu \Psi\, \partial^\mu \Psi + S_\text{CFT} + \int d^2x \, \Psi\, \mathcal{O} \,. 
\end{equation} 
The combined system is initialized in the product state $\left(\ketbra{0}\right)_\Psi \otimes \left(\rho_\beta\right)^\text{CFT}$, and evolved with the joint Hamiltonian deduced from the action above. Integrating out the holographic environment, we end up with the effective action for the open $\Psi$ system, which takes the form
\begin{equation}\label{eq:openM}
S_{open}[\Psi] = - \int d^2x\, 
\left(
   \partial_\mu \Psi_\skR\, \partial^\mu \Psi\skR
   - \partial_\mu \Psi_\skL\, \partial^\mu \Psi\skL
   \right)  + S_\text{gen}[\Psi_{\text{F}}, \Psi_\text{P}] \,.
\end{equation} 
The factorized kinetic term encodes the bare part of the system action, but the influence functions are obtained by replacing the sources for $\mathcal{O}$ by the corresponding retarded and advanced combinations of the field $\Psi$. One can read off from this action the effective couplings and deduce a classical stochastic model for the open system along the lines described in \cite{Jana:2020vyx}.

\subsection{Non-Markovian self-interactions}
\label{sec:nMself}

Let us now turn to the situation where we have a single field $\psi$, whose dual operator $\mathcal{P}$ has a long-lived mode. Within the context of our models of \cref{sec:BTZtoy}, such long-lived modes have diffusive dynamics. In higher dimensions, not only do we have diffusive dynamics, but also attenuated phonon modes. For the latter, the dilatonic modulation is more complicated. 

For simplicity, we will focus on a particularly simple example of a massless field, with Markovianity index $\ann <-1$ quantized with Neumann boundary conditions, since this is the situation that arises naturally in higher dimensional examples. The bulk dynamics is characterized by a single cubic coupling, 
\begin{equation}\label{eq:nMarkov3dmodel}
\begin{split}
S[\psi] 
&= 
   -\oint d^2x\,d\ctor\, \sqrt{-g}  
    \left[ 
   \frac{1}{2} \,\left(\frac{r}{r_+}\right)^{\abs{\ann}+1} \nabla_A \psi \nabla^A\psi    +\lambda\, \psi^3\right]  - \int d^2x\, \pi_\psi\, \psi\,.
\end{split}
\end{equation} 
We have indicated the explicit Neumann boundary term necessary to compute the generating function of correlators. Now the asymptotic fall-offs are $r^0$ and $r^{\abs{\ann}-1}$, with the latter defining the non-normalizable mode corresponding to the source $\widehat{J}$ for $\mathcal{P}$. The operator $\mathcal{P}$ has dimension $\tilde{\Delta} =0$ from the faster fall-off mode.
With this data we can write down the generating functional $S[\JFn,\JPn] $. This takes the same structural form as for the Markovian case \eqref{eq:MarkovS234}, which schematically we will write as $S[\JFn,\JPn]$. We define this with a different sign from the Dirichlet case to account for the Neumann boundary term. So 
\begin{equation}\label{eq:nMarkovS234}
S_\text{gen}[\JFn,\JPn] = S^{(2)}[\JFn,\JPn]  + S^{(3)}[\JFn,\JPn] + S^{(4)}[\JFn,\JPn] \,, 
\end{equation} 
where now
\begin{equation}\label{eq:nMS2}
\begin{split}
S^{(2)}[\JFn,\JPn] 
&=   \int_{k_1,k_2} \frac{\JPn(k_1)}{\nB(\omega_1)} \frac{\JFn(k_2)}{\Kin{\mathcal{P}}(k_2)} \delta(k_1+k_2)  \,,
\end{split}
\end{equation} 
with  
\begin{equation}\label{eq:KPdef}
\Kin{\mathcal{P}}(k) \equiv \frac{1}{\Kin{\ann,0}(k)} = 
-\frac{1}{(1-\abs{\ann})^2}\,  \Kin{-\abs{\ann},1-\abs{\ann}}(k)\,.
\end{equation} 
The non-Gaussian terms are given as before, cf., \eqref{eq:MarkovS234}.

While this generating functional has all the information one needs, owing to $\Kin{\mathcal{P}}(k)$ having gapless mode, the generating functional is non-local. These poles are explicit in the Gaussian term \eqref{eq:nMarkovS234}, but are also present in the non-Gaussian correlators, as can be discerned from our general discussion \cref{sec:analcorr} or directly read-off from \eqref{eq:Ipoles}. The origin of this non-local behaviour is easy to discern: in deriving $S[\JFn,\JPn] $ we have integrated out the low-lying non-Markovian quasinormal mode. The fix, as described in \cite{Ghosh:2020lel}, is obvious. We follow the Wilsonian logic, and retain the gapless mode in the low-energy description. One way to implement this is to Legendre transform the generating action to a Wilsonian influence functional parameterized by the expectation values of the operator $\mathcal{P}$. To wit, letting 
\begin{equation}\label{eq:Pexpval}
\snMar_\skR  \equiv \expval{\mathcal{P}_\skR} \,, \qquad \snMar_\skL  \equiv \expval{\mathcal{P}_\skL} \,,
\end{equation} 
we define\footnote{
   The coupling between sources and operators in the retarded-advanced (FP) basis follows directly from the couplings $\widehat{J}_\skR\, \mathcal{P}_\skR -\widehat{J}_\skL\, \mathcal{P}_\skL $ on the boundary Schwinger-Keldysh contour.
}
\begin{equation}\label{eq:nMWIF}
S_{_\text{WIF}}[\snMarF,\snMarP] = S_\text{gen}[\JFn,\JPn]  - \int_{k_1,k_2} \left( \frac{\JFn(k_1)\, \snMarP(k_2)}{\nB(\omega_2)} +  \frac{\snMarF(k_1)\, \JPn(k_2)}{\nB(\omega_2)} \right) \delta(k_1+k_2)
\end{equation} 

The Legendre transform is straightforward to carry-out. At leading order, we recover the expected relation between the sources and fields\footnote{
   In writing the expression, we have used the fact that overall 2-momentum reversal effectively is a frequency (or time) reversal in a system that is parity invariant.
} 
\begin{equation}\label{eq:PFnMsources0}
\JFn(k) = \Kin{\mathcal{P}}(k)\, \snMarF(k) \,, \qquad 
\JPn(k) = \Kin{\mathcal{P}}(\wk)\, \snMarP(k) \equiv  \Krev{\mathcal{P}}(k)\, \snMarP(k)\,.
\end{equation} 
In particular, note that the fields and sources are correctly related by the retarded and advanced sources (F and P, respectively).  This relation will get corrected perturbatively (in bulk coupling parameter $\lambda$) by the non-Gaussian terms. To get results to quartic order, it suffices to work out the correction from the cubic terms, viz., by solving the system 
\begin{equation}\label{eq:PFnMsources1}
\begin{split}
\frac{\JFn(-k)}{\Kin{\mathcal{P}}(-k)\, \nB(\omega)} 
+ \fdv{\JPn(k)}\bigg[\mathcal{I}_{_\text{FFP}}\cdot(\JF, \JF, \JP) + \mathcal{I}_{_\text{PPF}}\cdot(\JP, \JP, \JF) \bigg]
&= 
   \frac{\snMarF(-k)}{\nB(\omega)} \,,\\
\frac{\JPn(-k)}{\Kin{\mathcal{P}}(k)\, \nB(-\omega)} 
+ \fdv{\JFn(k)} \bigg[\mathcal{I}_{_\text{FFP}}\cdot(\JF, \JF, \JP) + \mathcal{I}_{_\text{PPF}}\cdot(\JP, \JP, \JF)\bigg]  
&= 
   \frac{\snMarP(-k)}{\nB(-\omega)}   \,, 
\end{split}
\end{equation} 
for the sources. 

The reader can deduce that the effect of Legendre transform at quadratic and cubic orders simply substitutes the classical relation between sources and fields obtained the Gaussian action \eqref{eq:PFnMsources0}. At quartic order, however, we have in addition a contribution from convolution of 3-point contributions arising from the correction captured by \eqref{eq:PFnMsources1}. We write the result somewhat schematically, but in a suggestive form using a replacement rule, as  
\begin{equation}\label{eq:nMarkovWIF234}
\begin{split}
S_{_\text{WIF}}[\snMarF,\snMarP] 
= 
   S_\text{gen}\bigg[\JFn \mapsto \Kin{\mathcal{P}}\,\snMarF\,, \JPn \mapsto \Krev{\mathcal{P}}\, \snMarP\bigg]  + \delta S^{(4)}_{_\text{WIF}}[\snMarF,\snMarP] \,.   
\end{split}
\end{equation} 
The last term corrects the quartic influence functional. Its effect is to ensure that the  Wilsonian influence functional has a well-behaved low energy expansion. 

One can see this as follows: while the cubic influence functions' analytic structure was governed solely by those of the boundary Green's function, the quartic influence functional had additional singularities from the intermediate factorization channels \eqref{eq:Ipoles}. First, we note that all the singularities in the external operator positions are removed by the leading part of the solution between the sources and fields \eqref{eq:PFnMsources0}. This is the rationale for writing the result using the replacement rule. 
To address the second set of singularities, realize that the contribution in $ \delta S^{(4)}_{_\text{WIF}}[\snMarF,\snMarP]$ is schematically  proportional to either $\mathcal{I}_{_\text{FFP}} \, \Kin{\mathcal{P}} \, \mathcal{I}_{_\text{PPF}}$, or $\mathcal{I}_{_\text{FFP}} \, \Krev{\mathcal{P}} \, \mathcal{I}_{_\text{PPF}}$. The specific structure is dictated by the factorization channel of the term under consideration. This term has the same set of singularities at the external operator insertions, but also has singularities in the intermediate channels. These intermediate factorization singularities cancel between the two terms. As with the construction of 1PI actions in QFTs, this is  exactly what the Legendre transform is supposed to achieve. 

The overall structure can be discerned from how one would organize the Schwinger-Keldysh perturbation theory in the boundary.  The retarded Green's function $\Kin{\mathcal{P}}(k)$  acts as the `kinetic term' for the field variable $\snMar$ and one has cubic vertices set by $\mathcal{I}_{_\text{FFP}}$ and $\mathcal{I}_{_\text{PPF}}$.  The Gaussian part has been computed for a conserved $U(1)$ current, and for the energy-momentum tensor in arbitrary dimensions (in both neutral and charged plasmas). The expressions for the non-Gaussian terms can be written down in our toy model (though we have chosen not to  explicitly do so).  

Let us turn to the open quantum system of a scalar $\Psi$ coupled to a non-Markovian operator $\mathcal{P}$. The dynamics is specified as in \eqref{eq:PsiOact}, with the replacement $\mathcal{O} \leftarrow \mathcal{P}$. This time we write down the effective field theory as the field $\Psi$ coupled to a gapless field $\snMar$. The effective action takes the form
\begin{equation}\label{eq:opennM}
\begin{split}
S_{open}[\Psi,\snMar] 
&= 
   - \int d^2x\, 
   \left(
   \partial_\mu \Psi_\skR\, \partial^\mu \Psi\skR
   - \partial_\mu \Psi_\skL\, \partial^\mu \Psi\skL
   \right)  + S_{_\text{WIF}}[\snMarF, \snMarP] \\
&\qquad 
   + \int_{k_1,k_2} \left( \frac{\Psi_\text{F}(k_1)\, \snMarP(k_2)}{\nB(\omega_2)} +  \frac{\snMarF(k_1)\, \Psi_\text{P}(k_2)}{\nB(\omega_2)} \right) \delta(k_1+k_2)
\end{split}
\end{equation} 
The coupling in the second line inverts back the Legendre transform. If we carry it out, we end up imprinting the long-lived dynamics of $\mathcal{P}$ into non-local terms in the open effective field theory of $\Psi$ alone. However, going by Wilsonian intuition, it is more natural to leave the result in the form given in \eqref{eq:opennM}, which is manifestly local, and admits a sensible low energy expansion.

\subsection{Interaction of Markovian and non-Markovian modes}
\label{sec:MnMinteract}

Let us finally turn to the case where we have two fields, one with a Markovian mode ($\phi$) and another with a non-Markovian mode ($\psi$). We can have two types of cubic interactions between the $\phi$ and $\psi$, so the bulk dynamics can be modeled as 
\begin{equation}\label{eq:mixedmodel}
\begin{split}
S[\phi,\psi] 
&= 
   - \frac{1}{2}\, \oint d^2x\,d\ctor\, \sqrt{-g}  
    \left[ 
   \,\left(\frac{r_+}{r}\right)^{\ann-1}\left(\nabla_A \phi \nabla^A\phi + m^2 \, \phi^2\right) + \left(\frac{r}{r_+}\right)^{\abs{\ann}+1} \nabla_A \psi \nabla^A\psi  \right] \\
&\qquad 
    -\oint d^2x\,d\ctor\, \sqrt{-g} \left[\lambda_1\, \phi^2 \, \psi + \lambda_2\, \phi\, \psi^2 \right]- \int d^2x\, \pi_\psi\, \psi \,.
\end{split}
\end{equation} 
The fields $\phi$ and $\psi$ are also taken to the of the form introduced in \cref{sec:Mself} and \cref{sec:nMself}, respectively. We, however, have switched off the self-interactions of the fields for simplicity, to focus on the physics of the interaction between short and long-lived modes. 

It is once again straightforward to obtain down the generating function for the correlators of the operators $\mathcal{O}$ and $\mathcal{P}$, dual to $\phi$ and $\psi$, respectively, in terms of the sources $J$ and $\widehat{J}$, viz., 
$S_\text{gen}[\JF,\JP,\JFn,\JPn]$, as delineated above. We want to Legendre transform this data to $S_{_\text{WIF}}[\JF,\JP, \snMarF, \snMarP]$ and obtain the influence functional parameterized by the expectation values of the non-Markovian field. We now describe the salient features for the two types of cubic couplings in turn.

Consider first the case where $\lambda_1 \neq 0, \lambda_2 =0$. We have a non-vanishing $\expval{\mathcal{O}\,\mathcal{O}\,\mathcal{P}}$ 3-point function, while the non-vanishing 4-point functions are 
$\expval{\mathcal{O}\,\mathcal{O}\,\mathcal{O}\,\mathcal{O}}$ and $\expval{\mathcal{O}\,\mathcal{O}\,\mathcal{P}\,\mathcal{P}}$. This implies that the generating functional only has terms like $\mathcal{I}_{_{\text{FF}\hat{\text{P}}}}$, $\mathcal{I}_{_{\text{PF}\hat{\text{F}}}}$, etc. Since the dependence on the non-Markovian sources is at most linear at cubic order, we can directly solve for them in terms of the non-Markovian field expectation  values and Markovian sources. We obtain 
\begin{equation}\label{eq:PFnMsources2}
\begin{split}
\JFn(k)
&= 
   \Kin{\mathcal{P}}(k)\,   \snMarF(k) 
   - \nB(\omega)\, \Kin{\mathcal{P}}(k)\,
   \fdv{\JPn(-k)}\bigg[\mathcal{I}_{_{\text{FF}\hat{\text{P}}}} \cdot(\JF, \JF, \JPn) + \mathcal{I}_{_{\text{FP}\hat{\text{P}}}}\cdot(\JF, \JP, \JPn) \bigg]
   \,,\\
\JPn(k)
&= 
   \Krev{\mathcal{P}}(k)\,   \snMarP(k) 
   - \nB(\omega)\, \Krev{\mathcal{P}}(k)\,
   \fdv{\JFn(-k)}\bigg[\mathcal{I}_{_{\text{PP}\hat{\text{F}}}} \cdot(\JP, \JP, \JFn) + \mathcal{I}_{_{\text{PF}\hat{\text{F}}}}\cdot(\JP, \JF, \JPn) \bigg]
   \,.\\
\end{split}
\end{equation} 

In the Wilsonian influence functional, the quadratic and cubic terms are again obtained by substituting \eqref{eq:PFnMsources0}. This also holds for the quartic couplings which mix the two fields, i.e., terms like $\mathcal{I}_{_{\text{PP}\hat{\text{F}}\hat{\text{F}}}}$. However, the purely Markovian influence functionals acquire a correction from the cubic pieces in \eqref{eq:PFnMsources2}. Let us again write the  Wilsonian influence functional reads
\begin{equation}\label{eq:WIFnMM}
\begin{split}
S_{_\text{WIF}}[\JF,\JP, \snMarF, \snMarP] 
&= 
    S_\text{gen}\bigg[\JF,\JP,\JFn \mapsto \Kin{\mathcal{P}}\,\snMarF\,, \JPn \mapsto \Krev{\mathcal{P}}\, \snMarP\bigg]  + \delta S_{_\text{WIF}}^{(4)} \,.
\end{split}
\end{equation} 
Then, for instance, the Markovian four point function 
$\mathcal{I}_{_\text{FFFP}}$ gets corrected by a term of the form:
\small
\begin{equation}\label{eq:deltaFFFP}
\delta\mathcal{I}_{_\text{FFFP}} 
=  
     - \int_k 
    \left[\fdv{\JPn(k)}\mathcal{I}_{_{\text{FF}\hat{\text{P}}}}\cdot(\JF, \JF, \JPn)\right]
    \Kin{\mathcal{P}}(k)
    \left[\fdv{\JFn(-k)}\mathcal{I}_{_{\text{FP}\hat{\text{F}}}}\cdot (\JF, \JP, \JPn)\right] \,.
\end{equation} 
\normalsize
This is again easy to intuit from the presence of only an FP propagator for the field $\snMar$. The corresponding change in the 4-point Markovian correlator $\mathcal{I}_{_\text{FFPP}}$ arises similarly from the FP channel.

A similar exercise can be carried out for the case where the coupling $\lambda_2\neq 0$. One just has to account for the corrections to the quartic terms arising from the Legendre transform. The cubic couplings correct the $\mathcal{I}_{_{\text{FF}\hat{\text{P}}\hat{\text{P}}}}$ type correlators. 

With this understanding it is again easy to write down the effective open quantum description for a field $\Psi$ coupled to the holographic environment. In this case we model the system-environment action as
\begin{equation}\label{eq:PsiOPact}
S_\text{SE} = - \int d^2x\, \partial_\mu \Psi\, \partial^\mu \Psi + S_\text{CFT} + \int d^2x \, \left[\kappa_1\, \Psi\, \mathcal{O} + \kappa_2\, \Psi\, \mathcal{P}\right] \,. 
\end{equation} 
This kind of coupling of a single field to both the Markovian and non-Markovian operators can arise if we couple our system to a conserved current operator of a holographic CFT. For example, the coupling $\partial_\mu \Psi\, J^\mu$ to a conserved $U(1)$ current $J^\mu$ is of this form, with the transverse photons being Markovian, and the longitudinal modes being diffusive. Similar statements hold for coupling to the energy-momentum tensor.  

We can write down following the preceding discussion the effective action for the open system. We integrate out the Markovian field $\mathcal{O}$, treating $\Psi$ as its source, whilst retaining $\snMar$ the field  parameterizing the expectation value of $\mathcal{P}$. One ends up with
\begin{equation}\label{eq:SopenMixed}
\begin{split}
S_{open}[\Psi,\snMar] 
&= 
   - \int d^2x\, 
   \left(
   \partial_\mu \Psi_\skR\, \partial^\mu \Psi\skR
   - \partial_\mu \Psi_\skL\, \partial^\mu \Psi\skL
   \right)  + S_{_\text{WIF}}[\kappa_1 \,\Psi_\skR, \kappa_1 \,\Psi_\skL, \snMarF, \snMarP] \\
&\qquad 
   + \int_{k_1,k_2} \kappa_2\left( \frac{\Psi_\text{F}(k_1)\, \snMarP(k_2)}{\nB(\omega_2)} +  \frac{\snMarF(k_1)\, \Psi_\text{P}(k_2)}{\nB(\omega_2)} \right) \delta(k_1+k_2) \,.
\end{split}
\end{equation} 
This completes our discussion of the open effective theory for a generic coupling to both Markovian and non-Markovian fields of the environment.

\section{Discussion}
\label{sec:discuss}

We have described a broad class of models for studying open quantum field theories, both with and without long-lived gapless modes. Our construction was  broadly inspired by earlier holographic analysis, in particular, the ability to model long-lived modes using designer scalars. While realistic examples have addition features, such as a more complicated radial mass or potential term, the essential point is the overall simplicity afforded by the holographic constructions.  

The open effective field theory is governed by real-time thermal correlators of the holographic environment. To compute these, we  exploit proposal of \cite{Glorioso:2018mmw}. In the process we strengthen and amplify a point made in \cite{Jana:2020vyx}, viz., that the grSK geometry provides the natural background for computing higher point functions using Witten diagrams.  In the present work we have explained how to compute exchange diagrams. These, in principle, could be exploited to also compute bulk loop effects. A useful corollary of our analysis is that thermal $n$-point functions are computed as radial integrals in a single copy of the exterior region of the stationary black hole spacetime. Moreover, the only data necessary is that of the ingoing boundary-bulk propagator. The integrands for the $n$-point functions are obtained as multiple discontinuity of a function built from them, and a  radial extension of the Boltzmann weight, the ubiquitous factors $e^{\beta\omega\zeta}$, which capture the monodromy picked up as we cross the horizon. These features can also be interpreted in terms of a `bulk open effective field theory' as will be discussed in \cite{Loganayagam:2022xyz}.

As noted in the main text some of these aspects have been touched upon in the literature earlier. For instance, \cite{Arnold:2011hp} discussed the computation of exchange diagrams in the black hole background. Their analysis relied on using advanced/retarded propagators in the bulk, but by working directly with our bulk-bulk propagator we have established that the grSK geometry respects all the Schwinger-Keldysh and KMS conditions. 

Our analysis here also touches upon several themes that have been discussed in the literature in related contexts. We will outline some lessons and open questions in these contexts, organized thematically, below. 

\paragraph{The grSK contour \& horizon localization:} As we have shown, the computation of bulk contact and exchange diagrams in the grSK geometry, reduces to the computation of a radial integral in the domain $r \in [r_+,\infty)$. The integrand for the contact diagrams is a single discontinuity of a combination of ingoing boundary-bulk propagators, radial Boltzmann weights, and vertex factors. A similar structure pertains to the exchanges, though now we have to take a multiple discontinuity. The ingoing propagators and $e^{\beta\omega\zeta}$ factors are regular at the horizon. Therefore, as noted below \eqref{eqn:Icontact} and in \cref{sec:analcorr}, as long as the vertex factors are analytic at the horizon, there is no localized contribution in the grSK geometry.  

In general for non-derivative polynomial interaction of fields, it would be non-covariant to have bulk vertices with explicit factors of $1/f$. Hence, horizon localized contributions are  precluded in such situations. However, as noted in \cref{fn:derint} these can occur when we have derivative couplings. This is for instance the case for fluctuations of a probe string studied in the context of Brownian motion in \cite{Chakrabarty:2019aeu}. There localized contributions arose owing to the derivative interactions from the Nambu-Goto action (leading to a double pole at the horizon).

Our interest in such localized contributions stems from \cite{Shenker:2014cwa}, who were analyzing 4-point out-of-time-order correlators and encountered such in the eikonal limit. While their analysis is a-priori not directly related to ours, it is interesting to inquire whether the there are situations in the grSK geometry with horizon localized contributions, and what they imply for boundary observables.

\paragraph{Hydrodynamic correlators:} In \cref{sec:BTZtoy} we pointed out the similarity between the two-dimensional toy model we analyzed, and a three-dimensional model with broken translational symmetry. As noted there \cite{Davison:2014lua} analyzed the behaviour of a probe Maxwell field and computed the two-point function of the boundary global current. Since the bulk Maxwell theory is Gaussian, this is the only observable. However, a variant of the model, an abelian-Higgs system, with a charged scalar field (also a probe), offers interesting opportunities to examine the interaction between fields that different infra-red behaviour. The Maxwell field has a non-Markovian piece in the charge density mode, which interacts now with the charged scalar field, along the lines sketched in \cref{sec:MnMinteract}.

Another context where there is a natural interaction between a Markovian and non-Markovian mode, is when a spinless primary field interacts with the energy density operator. In the gravitational dual, this maps to the interaction between the scalar field dual to the spin-0 primary interacts with the scalar polarizations of the gravitons. There are some interesting aspects to this, owing in particular to the fact that the energy density mode has a momentum dependent modulation. We hope to report on this in the near future as it appears to have a useful lesson for horizon localized contributions.

\paragraph{Relation to thermal bootstrap:}  As noted at the end of \cref{sec:Mself}, an interesting corollary of our analysis is the computation of thermal  four-point functions for scalar primaries of a 2d CFT. While their analytic structure is clear from the general discussion of \cref{sec:analcorr}, it would be useful to decompose the result in terms of thermal blocks on the cylinder and obtain an inversion formula along the lines of \cite{Caron-Huot:2017vep} for the thermal OPE data. The conformal bootstrap at finite temperature was broadly analyzed in \cite{Iliesiu:2018fao}. In the specific case of two-dimensional Euclidean CFTs \cite{Gobeil:2018fzy} examined torus conformal blocks. The aim here would be to set up the corresponding problem in real-time, using the momentum space data obtained for the influence functionals. In this context, it is also worth noting that thermal CFTs in momentum space  were analyzed in general dimensions in \cite{Manenti:2019wxs}, who also obtained the spectral density from the (cylinder) conformal blocks in two-dimensions.\footnote{
   It is also worth noting that retarded thermal two-point functions  can be used to deduce the high-spin asymptotics of heavy-light OPE as described in \cite{Dodelson:2022eiz} and obtained from the exact answer for 4d CFTs in \cite{Dodelson:2022yvn}. 
}

\paragraph{Comments on apparent quasinormal modes:}  Our analysis gives a clear picture of the  analytic structure for the boundary correlators. In the $\mathbb{C}^2$ parameterized by $\omega, \abs{\vb{k}}$ we encounter a codimension-1 loci of quasinormal spectral curves (labeled by a non-negative integer $n$). However, at discrete codimension-2 points, viz., at special kinematics, the correlators are analytic. These we decided to refer to as apparent quasinormal poles. By virtue of the arguments in \cref{sec:analcorr}, the same behaviour holds for the bulk Green's functions $\Gin{}$, $\Gout{}$, and  $\Gbulk{}$.

This observation helped us disambiguate some statements and terminology which we have found confusing in the literature, as noted in \cref{fn:nopolesareskipped}.  In attempting to find a relation between the physics of scrambling and many-body chaos, which is encoded in out-of-time-ordered four-point functions, and thermal energy density two-point functions, \cite{Blake:2017ris} argued for the phenomenon of `pole-skipping'. The phrase refers to (potential) poles of the retarded two-point functions, which are skipped because of an accidental zero. The particular locus nevertheless is claimed to control scrambling features of the higher-point correlator.

As one might expect, there is a story from the bulk analysis, which is revelatory. Initially,  \cite{Grozdanov:2017ajz} observed in the wave equation governing  energy density correlators, special codimension-2 loci in $\mathbb{C}^2$, additional modes that are normalizable at infinity and analytic at the horizon. This was  point was  fully appreciated by \cite{Blake:2018leo} who did a careful analysis of the wave equation, but as pointed out in \cite{Grozdanov:2019uhi,Blake:2019otz} the phenomenon is quite generic. 

Modes that are normalizable at the boundary do not however register in the boundary-bulk propagators. Moreover, no poles are technically skipped in the boundary Green's function \eqref{eq:Dqnms}. The function $\Kin{\ann,\Delta}(k)$ is meromorphic with simple poles at these loci (nb: meromorphic functions are rational functions). So if one was to examine, say the Mittag-Leffler form of the thermal two-point function, one would find no indication of a special values of frequency and momenta. One would simply read off the physical quasinormal modes where the correlator is singular. Consequently, there is no clear notion of pole-skipping if one examines just the retarded two-point functions alone. However, given the original motivation for the connection, this deserves further investigation.

\paragraph{Bulk loops and analytic structure:} Our analysis of non-Gaussian influence functionals was restricted to tree level Witten diagrams in the bulk. This captures the leading planar contribution to the correlation functions. Bulk loops, which can in principle be computed within our formalism, will give subleading corrections. It is interesting to ask whether the analytic structure found for the correlations, viz., that they are meromorphic with quasinormal (and anti-quasinormal) poles, is robust to such non-planar corrections.\footnote{
    We thank Shiraz Minwalla for discussions on this issue, and an anonymous referee for asking us to comment on it here.
} 
We speculate that in planar perturbation theory the analytic structure is not modified. 

To motivate this, consider the vacuum 2-point function, where the bulk 1-loop diagram gives the leading non-planar correction to the anomalous dimension. Focusing for simplicity on primaries  of dimension $\Delta$ in a 2d holographic CFT, the thermal 2-point function can be obtained from analytically continuing this result. In Fourier space, the resulting 1-loop answer takes the form quoted in \eqref{eq:Kindef}, albeit with a corrected conformal dimension,  $\Delta \to \Delta + \order{c^{-1}}$. Expanding out this in the anomalous dimension, we find the 1-loop answer can be expressed as a product of the tree level result and polygamma functions, which are meromorphic. While suggestive, it remains to confirm whether this expectation is borne out for $d>2$. One also ought to be able to address this without recourse to the analytic continuation directly within the grSK formalism.

\section*{Acknowledgements}

We would like to thank Veronika Hubeny and Godwin Martin for valuable discussions.
RL acknowledges support of the Department of Atomic Energy, Government of India, under project no. RTI4001, and would also like to acknowledge his debt to the people of India for their steady and generous support to research in the basic sciences. MR and JV would like to acknowledge the Aspen Center for Physics for hospitality during the course of this work, where they were supported in part by National Science Foundation Grant No. PHYS-1066293. 
MR would also like to thank ICTS, Bengaluru and TIFR, Mumbai for hospitality during the concluding stages of this project.  MR and JV were supported  by U.S.\ Department of Energy grant DE-SC0009999 and funds from the University of California. JV was supported by  U.S. Department of Energy grant DE-SC0020360 under the HEP-QIS QuantISED program (from September 2022).

\appendix

\section{Exchange diagrams on the grSK geometry}
\label{sec:exchangechecks}

In \cref{sec:Wittendiag} we provided the basic strategy for the computation of exchange diagrams, highlighting some consistency checks, which ensure that the grSK geometry does reproduce the expected thermal Schwinger-Keldysh correlators. In this appendix we provide some additional details of these checks, specifically for four and five-point functions. The analysis here can be iterated to higher point functions, but we shall not explicitly do so.

\subsection{Single bulk exchange diagrams}
\label{sec:4pt}

Four-point functions arising from a single exchange with cubic bulk vertices have been discussed in the main text. The generic structure of the integral we need to evaluate is given in \eqref{eq:F12integrals}. To get a sense for the integrands and to test the SK-KMS conditions, let us examine the cases where the boundary field theory correlators have to vanish. 

Consider first the case where all the external operators are of the F-type. We have four boundary-bulk propagators, each of which is ingoing (coefficient of $\JF$) and one bulk-bulk propagator. This results in the integral
\begin{equation}\label{eq:f4cf}
\begin{split}
\mathcal{I}_{_\text{FFFF}}^{a_1 a_2 a_3 a_4}(k_1,k_2,k_3,k_4) 
&= 
   \oint d\ctor\, \oint d\ctor' \;
    \Gin{a_1}(\ctor,k_1)\, \Gin{a_2}(\ctor,k_2)\,  \Gin{a_3}(\ctor',k_3)\,  \Gin{a_4}(\ctor,k_4)\, \Gbulk{e}(\ctor,\ctor';k) \,.
\end{split}
\end{equation}  
Writing out the bulk-bulk propagator \eqref{eq:Gbulk} we separate the two contour orderings as in \eqref{eq:4pt1exfull}. Each of these terms is separately required to vanish, and does so, owing simply to the periodicity property of the ingoing Green's function \eqref{eq:Ginperiod}. A similar argument holds for the PPPP correlator. Here the integrand is composed of the outgoing propagators for the external insertions, i.e., we replace $\Gin{a_i} \to \Gout{a_i}$ in the expression \eqref{eq:f4cf}. The outgoing propagator includes an additional exponential factor, which however conspire to cancel out upon using overall momentum conservation. The argument presented here for four-point correlators immediately generalizes to higher-point correlators with a single bulk exchange.

The vanishing condition is strongly contingent on the presence of the exponential factor of $e^{\beta\omega\,\ctor'}$ in the bulk-bulk propagator. It is easy to check that without it, the above integrands are non-vanishing. As we noted in the text, this factor is deduced by examining the Wronskian between the properly normalized basis of solutions. We have independently tried to constrain the bulk-bulk propagator directly by demanding the vanishing condition for all F or all P correlators (cf., \cref{fn:fixexpbb}). At the level of four-point functions, we find the constraints do not suffice fix to fix the functional dependence on the source point ($\ctor'$ above) completely. It however is possible that a similar exercise with a few higher point correlators could suffice. We will not undertake this exercise here, but argue below for consistency checks at the level of two bulk exchange processes. 

\subsection{Two bulk exchange diagrams}
\label{sec:5pt}

We now turn to the situation where we have two bulk exchanges on the grSK geometry. This is mostly to exemplify the general structure and provide further evidence for the consistency of our identification of the bulk-bulk propagator. 
For simplicity, we focus on five-point functions computed in a theory with cubic bulk vertices. 

With this assumption, topologically, the diagram is concatenation of two bulk-bulk propagators with five boundary-bulk propagators. Generally we have integrands of the form
\begin{equation}
\begin{split}
&
    \Gin{a_1}(\ctor,k_1) \,\Gin{a_2}(\ctor,k_2) \, \Gbulk{e_1}(\ctor,\ctor',k)\,\Gin{a_3}(\ctor',k_3) \,\Gbulk{e_2}(\ctor',\ctor'',k')\,\Gin{a_4}(\ctor'',k_4)  \,\Gin{a_5}(\ctor'',k_5) \\
&
    \Gin{a_1}(\ctor,k_1) \,\Gin{a_2}(\ctor,k_2) \, \Gbulk{e_1}(\ctor,\ctor',\bwt)\,\Gout{a_3}(\ctor',k_3) \,\Gbulk{e_2}(\ctor',\ctor'',\widetilde{\bwt})\,\Gout{a_4}(\ctor'',k_4)  \,\Gout{a_5}(\ctor'',k_5)  \,,
\end{split}
\end{equation}  
etc., where we have binary choices of $\Gin{\varphi}$ and $\Gout{\varphi}$ to attach to the sources $\JF$ and $\JP$, respectively.

Expanding out the bulk-bulk propagators in the above, we will end up with having to compute the following contour integral
\begin{equation}\label{eq:F1234integrals}
\begin{split}
I_\text{2-ex} 
&= 
   \oint d\ctor \oint d\ctor' \oint d\ctor'' \bigg[
   F_1 (\ctor,\ctor',\ctor'') \, \Theta(\ctor-\ctor')\, \Theta(\ctor'-\ctor'') + 
   F_2 (\ctor,\ctor',\ctor'') \, \Theta(\ctor-\ctor')\, \Theta(\ctor''-\ctor') \\
&\qquad \qquad + 
   F_3 (\ctor,\ctor',\ctor'') \, \Theta(\ctor'-\ctor)\, \Theta(\ctor'-\ctor'') 
   F_4 (\ctor,\ctor',\ctor'') \, \Theta(\ctor''-\ctor')\, \Theta(\ctor'-\ctor) \bigg]\,. 
\end{split}
\end{equation}  
For the vertex positions $\ctor, \ctor',\ctor''$  we have $3!$ orderings. But there are only $2^2 =4$ combinations of contour ordered theta functions arising from multiplying out the bulk-boundary propagators. The coefficient functions $F_1, \cdots F_4$ denoted above can be viewed as follows. The $F_1$ and $F_4$ terms are fully contour ordered (the former will be referred to as ordered, and the latter as anti-ordered), but the integrands $F_2$ and $F_3$ are only partially ordered. We use the contour step function identity \eqref{eq:stepidentity} to complete such partial orderings. This gives us  the six ordering expected for the 3 vertex integration positions. We then pick out as the following integrands:
\begin{equation}
\begin{split}
& F_1 (\ctor,\ctor',\ctor'') \, \Theta(\ctor-\ctor')\, \Theta(\ctor'-\ctor'') \\ 
& F_2 (\ctor,\ctor',\ctor'') \, \Theta(\ctor-\ctor')\, \Theta(\ctor''-\ctor') \left(\Theta(\ctor-\ctor'') + \Theta(\ctor'' -\ctor)\right) \\ 
& F_3 (\ctor,\ctor',\ctor'') \, \Theta(\ctor'-\ctor)\, \Theta(\ctor'-\ctor'') \left(\Theta(\ctor-\ctor'') + \Theta(\ctor'' -\ctor)\right)\\ 
& F_4 (\ctor,\ctor',\ctor'') \, \Theta(\ctor''-\ctor')\, \Theta(\ctor'-\ctor) \\ 
\end{split}
\end{equation}  
Our aim is to combine these using the contour ordering and reduce the result to an integral on a single-sheet, in analogy with \eqref{eq:4pt1exfull}.

To do so, we realize that we also have the freedom of placement of the operators in either leg of the bulk SK contour. Since the absolute ordering is fixed, we get only 4 choices for each of the six cases (altogether $24$ possibilities). Operationally, imagine fixing a permutation of the three vertices, say the fully ordered $\ctor > \ctor' > \ctor''$, and then side the positions consistent with the ordering on each of the sheets of the grSK contour cyclically  (i.e., from L to R through the horizon cap).  Finally, we decompose these, after placement of the vertices, correctly ordered, into single copy integrands, with the standard step functions. At the end of the day we find for each of the 6 single copy orderings eight possibilities, leading to altogether 48 terms. The count of 8 per single-copy collapsed order is easy to see as a triple discontinuity; each discontinuity/contour integral gives a pair of terms with relative sign.  

When the dust settles we find the following combination of terms for a two-exchange diagram
\begin{equation}\label{eq:5ptfinal}
\begin{split}
I_{2-ex} 
&= 
   \int_{\ctor_c}^{\ctor_+} d\ctor \, \int_{\ctor_c}^{\ctor_+} d\ctor' 
   \int_{\ctor_c}^{\ctor_+} d\ctor'' \bigg[
      \mathfrak{F}_1\, \theta(\ctor-\ctor')\, \theta(\ctor'-\ctor'') + 
      \mathfrak{F}_2\,  \theta(\ctor-\ctor'')\, \theta(\ctor''-\ctor') \\
&\qquad 
      +  \mathfrak{F}_3 \, \bigg( \theta(\ctor'-\ctor)\, \theta(\ctor-\ctor'') +
      \theta(\ctor'-\ctor'')\, \theta(\ctor''-\ctor)  \bigg)+ 
      \mathfrak{F}_4 \, \theta(\ctor''-\ctor)\, \theta(\ctor-\ctor') \\
&\qquad\qquad 
      + \mathfrak{F}_5\, \theta(\ctor''-\ctor')\, \theta(\ctor'-\ctor) 
\bigg] \,,
\end{split}
\end{equation} 
with 
\begin{equation}\label{eq:5ptFs}
\begin{split}
\mathfrak{F}_1    
&=
   F_1(\ctor,\ctor',\ctor'') - F_1(\ctor+1,\ctor',\ctor'') 
   + F_2(\ctor+1,\ctor', \ctor''+1) - F_2 (\ctor,\ctor',\ctor''+1)  \\
&\quad 
   + F_3(\ctor+1,\ctor'+1, \ctor'') - F_2 (\ctor,\ctor'+1,\ctor'') 
   + F_4(\ctor,\ctor'+1, \ctor''+1) - F_4 (\ctor+1,\ctor'+1,\ctor''+1) \,,\\
\mathfrak{F}_2
&=  
    F_2(\ctor,\ctor',\ctor'') - F_2(\ctor+1,\ctor',\ctor'') 
    + F_2 (\ctor+1,\ctor',\ctor''+1) -  F_2(\ctor,\ctor', \ctor''+1)  \\
&\quad 
   + F_3(\ctor,\ctor'+1, \ctor''+1) - F_3 (\ctor+1,\ctor'+1,\ctor''+1) 
   + F_3(\ctor+1,\ctor'+1, \ctor''+1) - F_3 (\ctor,\ctor'+1,\ctor'') \,,   \\
\mathfrak{F}_3     
&=
   F_1(\ctor+1,\ctor'+1,\ctor'') - F_1(\ctor+1,\ctor',\ctor'') 
    + F_2 (\ctor+1,\ctor',\ctor''+1) -  F_2(\ctor+1,\ctor'+1, \ctor''+1)  \\
&\quad 
   + F_3(\ctor,\ctor', \ctor'') - F_3 (\ctor,\ctor'+1,\ctor'') 
   + F_4(\ctor,\ctor'+1, \ctor''+1) - F_4 (\ctor,\ctor',\ctor''+1)  \,,   \\
\mathfrak{F}_4    
&=
    F_2(\ctor+1,\ctor',\ctor''+1) - F_2(\ctor+1,\ctor',\ctor'') 
    + F_2 (\ctor,\ctor',\ctor''+1) -  F_2(\ctor,\ctor', \ctor''+1)  \\
&\quad 
   + F_3(\ctor+1,\ctor'+1, \ctor'') - F_3 (\ctor,\ctor'+1,\ctor'') 
   + F_3(\ctor,\ctor'+1, \ctor''+1) - F_3 (\ctor+1,\ctor'+1,\ctor''+1)\,, \\    
\mathfrak{F}_5 
&=  
     F_1(\ctor+1,\ctor'+1,\ctor'') - F_1(\ctor+1,\ctor'+1,\ctor''+1) 
   + F_2(\ctor+1,\ctor', \ctor''+1) - F_2 (\ctor+1,\ctor',\ctor'')  \\
&\quad 
   + F_3(\ctor,\ctor'+1, \ctor''+1) - F_3 (\ctor,\ctor'+1,\ctor'') 
   + F_4(\ctor,\ctor', \ctor'') - F_2 (\ctor,\ctor'+1,\ctor''+1)   \,. 
\end{split}
\end{equation}
We have checked that these combinations of integrands vanishes for the FFFFF and PPPPP correlators. One can work out the corresponding combinations for the non-vanishing diagrams and express them in terms of a recursive master integral, along the lines sketched in \eqref{eq:recurseint}.


\providecommand{\href}[2]{#2}\begingroup\raggedright\endgroup

\end{document}